\documentstyle[preprint,aps,eqsecnum]{revtex}

\begin{document}
\def\subdecay#1{\raise 5pt \hbox{\hspace {0.1 em} {\vrule height 15pt
depth -2.8 pt } \hspace {-0.75 em} $\longrightarrow #1$} }

{\tighten
\preprint{\vbox{
\hbox{FERMILAB--PUB--96/104--T}}}

\title{\bf On the Necessity of Recalibrating Heavy Flavor Decays and
its Impact on Apparent Puzzles in High Energy Physics}
\author{Isard Dunietz}
\author{Internet: dunietz@fnth15.fnal.gov}

\address{\it Fermi National Accelerator Laboratory, P.O. Box 500,
Batavia, IL
60510}

\bigskip
\date{\today}
\maketitle
\begin{abstract}
It is demonstrated that charm is systematically undercounted in various
experiments. Via a process of elimination, $B(D^0 \rightarrow K^- \pi^+ )$ is
identified as the culprit. It calibrates essentially all charmed meson production
and decay properties, and thus is central to the physics of heavy flavors. We
predict it to decrease significantly below currently accepted values. We suggest
several novel methods for precise measurements of $B(D^0 \rightarrow K^- \pi^+ )$.
The $B(\Lambda_c \rightarrow pK^-\pi^+)$, on the other hand, calibrates heavy-flavored baryons. Its
world average relies heavily on a model of baryon production in $B$ decays, which
would be invalidated if $\overline B\rightarrow D^{(*)} \stackrel{(-)}{N} X$
processes were found to be significant. A Dalitz-plot analysis explains naturally
the soft inclusive $\Lambda_c$ momentum spectrum in $\overline B$ decays, and
predicts sizable $\overline B\rightarrow D^{(*)}\stackrel{(-)}{N} X$ processes.
Consistently carrying through these modifications to charmed meson and baryon yields has the potential to resolve
the heavy-flavor puzzles at $Z^0$-factories [$R_c,\; R_b$], the number of charm per $B$-decay
puzzle, and the semi-leptonic $B$ decay puzzles. We emphasize that state of the
art theoretical calculations are consistent with precise experimental measurements
regarding $B(\overline B\rightarrow X\ell\bar\nu )$. Recent CLEO measurements are
interpreted as newly available cross-checks that any inclusive theoretical
investigation must satisfy.
Another topic of this report concerns the $b\rightarrow c + \overline D^{(*)}\overline K$
transitions, which were predicted to be sizable and subsequently confirmed by CLEO. This report discusses the
underlying dynamics of those processes and quantifies the necessary modifications
in existing semileptonic analyses due to the ignored $b\rightarrow \overline
D\rightarrow \ell^-$ background. The determination of the average, time-integrated
$B-\overline B$ mixing parameter $\overline \chi$ at higher energy colliders is more
subtle than currently realized. All existing dilepton analyses that determine
$B(b\rightarrow \ell^- )$ must modify their removal of $B-\overline B$ mixing
effects. Several implications on the physics of heavy flavors are mentioned.
\end{abstract}
}

\newpage

\section{Introduction}

One of the purposes of this paper is to demonstrate that there is a
systematic undercounting of charm in various experiments.
Whereas the number of primary charm at $Z^0$ factories $(R_c)$ is
accurately
predicted to be 17.2\%, experiment observes only $(15.98 \pm 0.69)$\%
\cite{lepew96,graziani,charlton,opalrc}.
Historically, the number of charmed hadrons per $B$ decay $(n_c)$ was
measured to
be smaller than expectations~\cite{nchistory}, and the sum over all
exclusive semileptonic $B$
decay BR's is significantly less than the inclusive $B(B\rightarrow
X\ell\nu
)$ measurements \cite{drellp}.
A common thread in all of these cases is charm counting which
heavily depends on the value of the branching fraction for
$D^0\rightarrow K^- \pi^+$.

The value of $B(D^0 \rightarrow K^-
\pi^+ )$ calibrates much that is known in charm and beauty decays,
and is believed to be known to better than $\pm$ 5\% accuracy.
The experiment CLEO~\cite{cleod0} measures
$$ B(D^0 \rightarrow K^- \pi^+) = (3.91 \pm 0.19) \% \; ,$$
and the 1994 Particle Data Group~\cite{pdg} cites a world average of
$$ B(D^0 \rightarrow K^- \pi^+) = (4.01 \pm 0.14) \% \; .$$
These values have been used to calibrate not only most other
$D^0$ decay modes, but $D^+$ decay modes as well \cite{cleorp}, via
$$r_+ =\frac{B(D^+\rightarrow K^- \pi^+ \pi^+)}{B(D^0 \rightarrow K^-
\pi^+ )}
\;.$$
The calibration mode for the $D_s$, namely $B(D_s \rightarrow \phi\pi )$, has
also been recently tied to $B(D^0 \rightarrow K^- \pi^+ )$ in a
model independent fashion \cite{cleods}.

Historically $n_c$ has been obtained by combining all charmed hadron yields
in $B$ decays, including
the inclusive production of weakly decaying charmed baryons.
The present central values for $\Xi_c$ and $\Lambda_c$
yields which are used to calculate the baryon component of $n_c$
\cite{ncbrowder} are comparable at roughly 5\%.
The latter is measured rather
well, whereas the $\Xi_c$ measurement has large uncertainty.
The CLEO collaboration has demonstrated that the
right-sign $\ell^+\Lambda_c$ correlations dominate over the wrong-sign $\ell^-
\Lambda_c$ case \cite{glasgowbaryon} (where the lepton comes from the
semileptonic decay of one $B$ and the $\Lambda_c$ originates from the other
$B$ in an  $\Upsilon (4S)$ event).
 As a result, the inclusive $\Xi_c$ production
in $B$ decays cannot, in fact, be as large as that of the $\Lambda_c$.
[Throughout this report, CP violation is neglected
and for each process its CP-conjugated relative is implied.]

Because $\Lambda_c$ production in $B$ decays is measured
with greater accuracy than that of the $\Xi_c$, we correlate
the weakly decaying charmed baryon $(baryon_c)$ production in $B$ decays
to that measured for the $\Lambda_c$ (see Appendix). In
the process, we find that the current measured central
value for the $\Xi_c$ yield in $B$ decays is too high.
We also propose to not use the 1994 PDG value~\cite{pdg} of
$B(\Lambda_c\rightarrow pK^- \pi^+ )
= (4.4 \pm 0.6)$\%, because it relies heavily on a flawed model of
baryon production in $B$ decays. We instead follow the more satisfactory
approach outlined in Ref.~\cite{shipsey}, where
$B(\Lambda_c\rightarrow pK^- \pi^+)=(6.0 \pm 1.5)$\% is obtained by
assuming equal inclusive semileptonic widths for $D$ and $\Lambda_c$ decays.
These two effects reduce the $baryon_c$ production in $B$ decays
significantly and yield $n_c = 1.10 \pm 0.06$ when one
uses the value $ B(D^0 \rightarrow K^- \pi^+)~~=~~(3.91 \pm 0.19) \%.$
This is appreciably lower than current
$n_c$ estimates~\cite{ncbrowder}.

The CLEO \cite{moriond} experiment recently confirmed our prediction \cite{bdy}
for significant wrong-charm production $(\overline B\rightarrow
\overline DX)$ in $B$ decays and has therefore completed the mapping of the
$b\rightarrow \bar c$
transitions. This opens up an alternate, indirect method for the determination
of $n_c$ which is far less sensitive to either
$B(\stackrel{(-)}{B}\rightarrow baryon_c X)$ or
$B(D^0 \rightarrow K^- \pi^+ )$. This method yields a
value of $1.19 \pm 0.03$, which differs from the direct measurement.

Since inclusive charm production in $B$ decays is dominated by
the inclusive $D$ yields, we will entertain the possibility
that the current accepted value for $B(D^0 \rightarrow K^- \pi^+)$
could be wrong.  [We consider $B(D^0 \rightarrow K^- \pi^+)$
instead of the absolute branching fractions for $D_s$ and
$\Lambda_c$, because the latter play a smaller role in the
determination of $n_c$ and would therefore have to be altered
beyond reasonable extremes to have the same result.
In addition, the Standard Model allows a sufficiently accurate estimate
for hidden charmonia production in $B$ decays and for charmless
$B$ decays.  We are thus naturally led to focus on
the $B(D^0 \rightarrow K^- \pi^+)$ (see Eqs. (2.21)-(2.22)).]
Since the two determinations for $n_c$ must agree, we equate them and
solve for $B(D^0 \rightarrow K^- \pi^+ )$ to obtain the precise result
$B(D^0  \rightarrow K^- \pi^+ )=(3.50 \pm 0.21)\%$, which is significantly
smaller than the currently
accepted value. This reduced value will be seen to solve the $R_c$ puzzle
and to mitigate
the semileptonic $B$ decay problem mentioned above.

If correct,
a reduced value for $B(D^0 \rightarrow K^- \pi^+ )$ would have
implications for the whole field of heavy flavor decays including
the discrepancy between theory and experiment regarding $R_b$.
A downward shift in the value for $B(D^0\rightarrow K^- \pi^+ )$
could imply that current experimental analyses of $R_b$ are
underestimating their tag rate for $Z\rightarrow c\bar c$.
Because $D^0 \rightarrow K^- \pi^+$
calibrates almost all charmed meson branching fractions, a reduced
value would necessarily imply that a proper accounting has yet to be
carried out for a significant fraction of $D^0$ decays. The
modes which have been miscounted or simply missed would likely involve
higher track multiplicities, since such decays are more vulnerable to
detector inefficiencies, particle misidentification, and the presence of
neutral daughters \cite{appel}. On the other hand, the higher multiplicities
would also mean that these decays would more easily be tagged as $B$'s.
We have studied this and other sources that have the
potential to remove the $R_b$ discrepancy between the Standard Model
and experiment\cite{distw}. There are, of course, many additional
implications, some of which we will touch upon in the conclusion.

Because the precise knowledge of the value of
$B(D^0 \rightarrow K^- \pi^+)$ is so central to these issues, we
encourage a widespread effort to remeasure it, and suggest several novel
methods for doing so.  It is noteworthy that the most recent
determination of $B(D^0 \rightarrow K^- \pi^+)$, as obtained by
ARGUS~\cite{argusd0}, is compatible with our findings,
$$
B(D^0 \rightarrow K^- \pi^+) = (3.41 \pm 0.12 \pm 0.28)\% \;.
$$

The second topic of this note concerns $\overline B\rightarrow
D\overline D \;\overline KX$ processes. Because they have been
generally overlooked in published experimental analyses, it is of some
consequence that we predicted, in a variety of
ways, that they are sizable \cite{bdy}. CLEO \cite{moriond} confirmed our
prediction and has found that the inclusive wrong-charm yield is about 10\%,
$$B(\overline B\rightarrow \overline DX) \approx 10\%\;.$$
Since this comprises a large fraction of all $B$ decays, we will
discuss the underlying dynamics to help enable a more careful probing of these
transitions. We then shift gears and discuss
necessary modifications to published results for $B$ decays, which have not
considered the $\overline B\rightarrow
D\overline D \;\overline KX$ background. In the interest of
brevity we will give a detailed discussion only for measurements of
the semileptonic decays of $B$ hadrons.

We point out that current measurements~\cite{burchatr} of $\overline
B\rightarrow D^{**} (X)
\ell\bar\nu$ have to carefully assess the impact of the $b\rightarrow
c\; \overline
D\;\overline KX$ background, where one of the charmed hadrons
delivers the charged
lepton via its semileptonic decay.

The inclusive, single lepton spectrum determines
$B(\overline B\rightarrow X\ell\nu)$. CLEO subtracted leptons via
$\overline B\rightarrow (\Lambda_c ,D^-_s)\rightarrow \ell$
transitions \cite{wang,cleoslincl}. The shapes of the primary $(\overline
B\rightarrow
\ell^-)$ and secondary $(\overline B\rightarrow D\rightarrow \ell^+)$
spectra
were taken from various models. The overall normalizations for the
primary and
secondary spectra were obtained from a fit to data.
It was found that the ACCMM model~\cite{accmm} fits the data well,
whereas the ISGW model~\cite{isgw,isgw2} does not.
By letting the $\overline B\rightarrow D^{**} \ell\bar\nu$ component
float in the
ISGW model, CLEO~\cite{wang} could get a much better fit, albeit with
a very large value for the ratio;
$$\frac{B(\overline B\rightarrow D^{**} \ell\bar\nu)}{B(\overline
B\rightarrow X\ell\bar\nu )} = 0.23 \pm 0.01\;\pm 0.05\;.$$
CLEO denotes the modified ISGW model by ISGW$^{**}$.
All current analyses assume that inclusive $D$ production in
$\overline B$ decays is mediated by $b \to c$ transitions.  The
leptons generated from these $D$ decays have the opposite charge of the
primary leptons.
We now know that wrong-charm production in $\overline B$ decays is a
significant
source of soft $\overline D$'s~\cite{moriond}. The soft $\overline D$
gives rise to a soft lepton with the same charge as the primary lepton.
Of the two sources of secondary leptons ($b \to D \to \ell^+$ and $b \to
\overline D \to \ell^-$) the latter must be subtracted from the
primary component measured from charge
correlations~\cite{argusdilepton,cleoslincl,wang,alephslincl}. This
subtraction is lacking in all existing analyses.  The secondary
lepton yield $\overline B \to \stackrel{(-)} {D} \to \ell$ from
single inclusive lepton fits is enhanced relative to the $\overline B \to  D
\to \ell^+$ yield obtained from dilepton analyses.  We estimate the
enhancement due to $B(\overline B \to  \overline D \to e^-)$ to be about 1\%.
This systematic enhancement is indicated in Table 4.7 of
Ref.~\cite{wang}.

Once backgrounds have been treated correctly, the
inclusive single lepton analysis at $\Upsilon(4S)$ factories
\cite{wang} determines the correct primary lepton spectrum since
the method is insensitive to the charge of the lepton.  In contrast,
the published dilepton methods measure the primary lepton spectrum
for which the soft $\overline B \to  \overline D \to \ell^-$ background
contribution still must be removed.

Several analyses at the $Z^0$ resonance take the
primary lepton spectrum from $\Upsilon(4S)$ factories as input.
Those analyses ought to be updated, because they rely on older CLEO
data~\cite{henderson} with a softer primary lepton spectrum,
$$\frac{B(\overline B\rightarrow D^{**} \ell\bar\nu)}{B(\overline
B\rightarrow
X\ell\bar\nu )} = 0.32 \pm 0.05\;\;,$$
than recent CLEO results~\cite{wang}.  All inclusive $B(b\rightarrow
\ell^- )$ measurements at the $Z^0$ that rely on the primary spectral shape
from $\Upsilon(4S)$ factories will thus change.

The charge correlation method of ALEPH does not rely upon the spectral
shape from lower energy measurements. The basic idea for this analysis
originates from the ``model independent'' dilepton
analysis of ARGUS \cite{argusdilepton}, which was improved by CLEO
\cite{wang,cleoslincl}.
In such analyses, the primary and secondary components are
``model independently" extracted via charge correlations. However,
the analyses of
ARGUS, CLEO and ALEPH must be modified in two respects. First, because the
$b\rightarrow \overline D\rightarrow\ell^-$ background has not yet
been subtracted, the primary component is deceptively enhanced. Second,
the removal of $B^0 - \overline B^0$ mixing effects is more subtle than
currently realized.

Because most model dependence is indeed absent, we will be able to
quantify our expectations. The threshold machines operating at $\Upsilon (4S)$
require a momentum cut $p>0.6$ GeV/c for the signal electron. Fortunately, this
cut removes most of the soft
$\overline B\rightarrow \overline D\rightarrow e^-$  background
and drastically reduces the modification required for the correct removal
of $B_d -\overline B_d$ mixing effects. Experiments at the
$Z^0$ resonance are less fortunate. In this case, the large boost imparted
to the primary $B$ hadrons causes the background leptons to contribute more
significantly. As a consequence, the
model independent measurement of $B(b\rightarrow \ell^-)$ at the
$Z^0$ resonance will need to be shifted downward,
alleviating the puzzle of why the
inclusive semileptonic BR of $B$ hadrons at $Z^0$ factories is
significantly larger than at its
$\Upsilon (4S)$ counterpart \cite{browderh95,neubert}.

The model independent extraction of $B(b \to
\ell^-)$~\cite{alephslincl} is not the only observable that is affected by a
modification of the $B - \overline B$ mixing effects.
The published measurements~\cite{mixingwrong} of the average,
time integrated, mixing parameter $\overline \chi$ at higher energy
facilities must also be modified.  Whereas existing analyses have
implicitly assumed the same average mixing parameter $\overline \chi$
for the primary and secondary lepton components, we will demonstrate that
the two components experience different average $B - \overline B$
mixing effects.  Consequently, those
extractions~\cite{lepew96,alephslincl} of
$B(b \to \ell^-)$ that involve the average mixing parameter
$\overline \chi$ must be adjusted.

We hope to motivate our experimental colleagues to remeasure the quantities
of interest, since they alone can correctly assess all of the uncertainties.
We are eager to learn the outcome of their studies.

This report is organized as follows:

Section II determines $B(D^0 \rightarrow K^-\pi^+ )$
by equating the results for two complementary methods of determining $n_c$.
The resultant value of $B(D^0 \rightarrow K^-\pi^+ )$ is lower than the current
accepted values. Consequences of the theory of inclusive $B$ decays are
mentioned.

Section III interprets the low
$R_c$ measurements as another indication of a smaller than
expected value for $B(D^0 \rightarrow K^- \pi^+).$

Section IV discusses the underlying dynamics of $\overline
B\rightarrow D\overline
D\;\overline KX$ processes.

Section V reviews semileptonic $B$ decays. It first
discusses the structure of single, inclusive lepton data samples followed by
dilepton samples. Modifications resulting from the hitherto neglected
$b\rightarrow  \overline D\rightarrow\ell^-$ background, and
the correct removal of $B^0 -\overline B^0$
mixing effects are presented in great detail. Exclusive semileptonic
$B$ decays are then classified. Combining all classes of these
decays again illustrates the need for a lower value of
$B(D^0\rightarrow K^- \pi^+)$. Instead of using
inconclusive experimental $B(\overline B\rightarrow D^{**}
(X)\ell\bar\nu )$, we
suggest, for now, to use a Bjorken-Isgur-Wise sumrule~\cite{sumrule},
which ``model independently" relates the combined semileptonic BR into charmed
$p$ wave states to the Isgur-Wise slope parameter, $\rho^2$.

Section VI discusses the possibility of
novel precision measurements of $B(D^0 \rightarrow K^- \pi^+ )$ from
semileptonic $B$ decays. (Other new measurements were advertised in
earlier sections.)

Section VII explains why we prefer $B(\Lambda_c \rightarrow pK^- \pi^+ ) = (6.0 \pm 1.5)\%$
over the smaller world average $(4.4 \pm 0.6)\%$.

Section VIII discusses some of the consequences of a smaller than
currently accepted value for $B(D^0 \rightarrow K^-\pi^+)$. It also discusses
several implications of significant $b\rightarrow c + \overline
D^{(*)} \overline K$ transitions.

We believe that the adjustments to published, experimentally
extracted quantities suggested in this report are reasonable
estimates of what is required for a consistent picture
of heavy flavor decays. Nevertheless, we emphasize
the fact that the most accurate values and their systematic uncertainties
can only be obtained by the various experiments involved in these measurements
by means of the analysis of new data or reanalysis of existing data.
We are eager to learn from our experimental colleagues what results are
obtained after the effects discussed in this note have
been taken into account.

\section{$B(D^0 \rightarrow K^- \pi^+ )$ from $n_c$}

The number of charmed hadrons per $B$ decay is defined by
\begin{equation}
n_c \equiv Y_D +Y_{D_s} + Y_{baryon_c} + 2 B(\overline B\rightarrow
(c\bar c)X)
\;, \end{equation}
where the inclusive production of an arbitrary hadron $T$ is defined as
\begin{equation}
Y_T \equiv B(\overline B\rightarrow TX) + B(\overline B\rightarrow
\overline
TX) \;.
\end{equation}
Here all weakly decaying, singly charmed baryon species
$(\Lambda_c ,
\Xi^{+,0}_c , \Omega_c )$ are denoted collectively by $baryon_c$ while
$(c\bar c)$
denotes charmonia not seen as open charm.
Table I summarizes relevant CLEO measurements.
Note that the inclusive $D^+$ yield in
$B$ decays involves $B(D^+ \rightarrow K^- \pi^+ \pi^+)$, which in turn is
calibrated by $D^0 \rightarrow K^- \pi^+$ \cite{cleorp},
\begin{equation}
\frac{B(D^+ \rightarrow K^- \pi^+ \pi^+ )}{B(D^0 \rightarrow K^-
\pi^+ )} =
2.35 \pm 0.23 \;.
\end{equation}
We can thus write
\begin{eqnarray}
Y_{D^+} & = & (0.235 \pm 0.017) \;\frac{9.3}{3.91 (2.35 \pm 0.23)}
\;\left[\frac{3.91\%}{B(D^0 \rightarrow K^- \pi^+ )}\right] =
\nonumber \\
& = & (0.238 \pm 0.029) \;\left[\frac{3.91\%}{B(D^0 \rightarrow K^-
\pi^+
)}\right] \;.
\end{eqnarray}
The inclusive $D$ yield in $\overline B$ decays,
\begin{equation}
Y_D \equiv Y_{D^0} + Y_{D^+} \;,
\end{equation}
can then be expressed as shown in Table I.
The measurement of inclusive $\Xi_c$ production in $B$ decays
has large uncertainty because it suffers from a low number of candidate
events and a large uncertainty in the branching fraction used for its
calibration. In the Appendix, we therefore
correlate both $\Xi_c$ and $\Omega_c$ production in
tagged and untagged $B$ decays to $Y_{\Lambda_c}$ and
\begin{equation}
r_{\Lambda_c} \equiv \frac{B(\overline B\rightarrow
\overline\Lambda_c
X)}{B(\overline B\rightarrow \Lambda_c X)}\;.
\end{equation}
We neglect $b\rightarrow u$ transitions and use the Cabibbo suppression
factor, $\theta^2 =(0.22)^2$, for charmed baryon production in
$b\rightarrow c\bar us (b\rightarrow c\bar cd)$ versus $b\rightarrow
c\bar ud^\prime
(b\rightarrow c\bar cs^\prime)$.  [The prime indicates that the
corresponding Cabibbo suppressed mode is included.]  The Appendix
also parametrizes $s\bar s$ fragmentation from the vacuum, and predicts
\begin{equation}
\frac{Y_{\Xi_c}}{Y_{\Lambda_c}} = 0.38 \pm 0.10 \;,
\end{equation}

\begin{equation}
\frac{Y_{baryon_c}}{Y_{\Lambda_c}}= 1.41 \pm 0.12 \;,
\end{equation}

\begin{equation}
\frac{B(\overline B\rightarrow baryon_c X)}{Y_{\Lambda_c}} = 1.22 \pm
0.07 \;,
\end{equation}

\begin{equation}
\frac{B(\overline B \rightarrow \overline{baryon_c}
\;X)}{Y_{\Lambda_c}} = 0.20
\pm 0.10 \;.
\end{equation}
We probably overestimate $\Xi_c$ production in $\overline B$
decay, because our unsophisticated
predictions do not include the fact that the V-A interactions tend
to create highly excited $\Xi_c$ baryons in $\overline B$ decays (see the
Appendix).  The quark flavor of an initially highly excited $\Xi_c$
baryon is generally not retained by its weakly decaying offspring.
Nevertheless, the true value for $Y_{baryon_c} /
Y_{\Lambda_c}$ must lie in the range
\begin{equation}
1 < \frac{Y_{baryon_c}}{Y_{\Lambda_c}} < 1.41 \pm 0.12 \;.
\end{equation}
Variation over this range has negligible effect upon
the accurate extraction of $n_c$, and so,
for now, we use the values Eqs. (2.7) - (2.10), and hope to return
with a more realistic modelling of charmed as well as
uncharmed baryon production in $B$ decays in a future report.
Thus $n_c$ becomes:
\begin{eqnarray}
n_c & = & (0.883 \pm 0.038) \;\left[\frac{3.91\%}{B(D^0 \rightarrow
K^- \pi^+
)}\right] + (0.1211 \pm 0.0096) \;\left[\frac{3.5\%}{B(D_s
\rightarrow \phi\pi
)}\right] + \nonumber \\
& + & (0.042 \pm 0.008) \;\left[\frac{6\%}{B(\Lambda_c \rightarrow
pK^- \pi^+
)}\right] + 2B(\overline B\rightarrow (c\bar c) X) \;.
\end{eqnarray}
Inserting the necessary absolute branching ratios (BR) for charm decays (see
Table II) and estimating \cite{bdy}
\begin{equation}
B(\overline B\rightarrow (c\bar c)X) = 0.026 \pm 0.004 \;,
\end{equation}
one would obtain

\begin{equation}
n_c = 1.10 \pm 0.06\;\;\;\;,
\end{equation}
which is below the currently accepted value~\cite{ncbrowder}.

Most recently, CLEO completed the direct measurement of
$B(b\rightarrow c\bar
cs^\prime )$ which allows an alternative extraction of $n_c$ via the expression
\cite{bdy},
\begin{equation}
\tilde{n}_c = 1- B(b\rightarrow \;{\rm no}\;{\rm charm}) +
B(b\rightarrow c\bar
cs^\prime ) \;.
\end{equation}
This alternative extraction of $n_c$ is far less sensitive to
miscalibrations of absolute branching ratios of charmed hadron decays.
For $B(b\rightarrow$no charm) we use \cite{bdy},
\begin{eqnarray}
B(b\rightarrow \;{\rm no}\;{\rm charm})
& = & (0.25 \pm 0.10) \;(0.1049 \pm
0.0046) = \nonumber \\
& = & 0.026 \pm 0.010
\end{eqnarray}
The inclusive wrong-charm $B$ decay yield is \cite{fwd,bsbsbar,bdy,bucbars}
\begin{eqnarray}
B(b\rightarrow c\bar cs^\prime ) & \approx & B(\overline B\rightarrow
\overline D X)
+ B(\overline B\rightarrow D^-_s X) + \nonumber \\
& + & B(\overline B\rightarrow \overline{baryon}_c \;X) + B(\overline
B\rightarrow (c\bar c) X) \;.
\end{eqnarray}
From Tables I and III, Eq.~(2.17) and our charmed baryon model, we find
\begin{eqnarray}
B(b\rightarrow c\bar cs^\prime ) & = & (0.085 \pm 0.025)
\;\frac{3.91\%}{B(D^0
\rightarrow K^- \pi^+ )} + \nonumber \\
& + & (0.100 \pm 0.012) \;\left[\frac{3.5\%}{B(D_s \rightarrow
\phi\pi
)}\right] +(0.0059 \pm 0.0031) \;\left[\frac{6\%}{B(\Lambda_c
\rightarrow pK^-
\pi^+ )}\right] + \nonumber \\
& + & B(\overline B\rightarrow (c\bar c) X) \;.
\end{eqnarray}
With the absolute charm branching values from Table II we obtain
\begin{equation}
B(b\rightarrow c\bar cs^\prime )  = 0.22 \pm 0.03 \; ,
\end{equation}
\begin{equation}
\tilde{n}_c = 1.19 \pm 0.03 \;.
\end{equation}
Since $n_c$ and $\tilde{n}_c$
must be equal, the apparent discrepancy between them (compare Eqs.
(2.14) with (2.20)) indicates to us a possible miscalibration of absolute
BR's of charmed hadron decays. We hypothesize that the
problem is limited to $B(D^0 \rightarrow K^- \pi^+ )$ (see Section I) and
propose an accurate method for determining
this quantity as follows.
By treating $B(D^0 \rightarrow K^- \pi^+ )$ as an unknown and
equating $n_c$ with $\tilde{n}_c$, one gets
\begin{eqnarray}
B(\overline B\rightarrow DX) & = & 1- B(\overline B\rightarrow \;{\rm
no}\;{\rm charm}) -
B(\overline B\rightarrow D^+_s X) + \nonumber \\
& - & B(\overline B\rightarrow baryon_c X) - B(\overline B\rightarrow
(c\bar c ) X) \;.
\end{eqnarray}
Eq.~(2.21) expresses the trivial fact that \cite{bucbars}
$$B(b\rightarrow c)\approx 1-B(b\rightarrow {\rm no}\;{\rm charm}) \;,$$
and yields\begin{equation}
B(D^0 \rightarrow K^- \pi^+) =
\frac{3.91\% \cdot \left[B\left(\overline B\rightarrow
DX\right)\right]
|_{{\rm for} \;B(D^0 \rightarrow K^- \pi^+ ) = 3.91\%}}
{1-B(\overline B \rightarrow {\rm no}\;{\rm charm}) - B(\overline
B\rightarrow D^+_s X) -
B(\overline B\rightarrow baryon_c X) - B(\overline B\rightarrow
(c\bar c) X)} \;.
\end{equation}
Inserting the current CLEO data
\begin{equation}
B(\overline B\rightarrow DX) = (0.798 \pm 0.042) \left[\frac{3.91\%
}{B(D^0 \rightarrow
K^- \pi^+ )}\right] \;,
\end{equation}

\begin{equation}
B(\overline B\rightarrow D^+_s X) = (0.021 \pm 0.010)
\left[\frac{3.5\%}{B(D_s \rightarrow
\phi\pi )}\right] \;,
\end{equation}

\begin{equation}
B(\overline B\rightarrow baryon_c X) = (0.0365 \pm 0.0065
)\left[\frac{6\%}{B(\Lambda_c
\rightarrow pK^- \pi^+ )}\right] \;,
\end{equation}
and Eqs.~(2.13) and (2.16) we obtain
\begin{equation}
B(D^0 \rightarrow K^- \pi^+) = (3.50 \pm 0.21)\% \;.
\end{equation}
This in turn yields
\begin{equation}
B(b\rightarrow c\bar cs^\prime )  = 0.227 \pm 0.035 \; ,
\end{equation}
\begin{equation}
n_c = \tilde{n}_c = 1.20 \pm 0.04 \;.
\end{equation}

Once the model independent measurement invented by CLEO for
\cite{cleods}
$$\frac{B(D_s\rightarrow \phi\pi )}{B(D^0 \rightarrow K^- \pi^+ )}$$
becomes accurate enough, we suggest that it be used to determine
$B(D^0 \rightarrow  K^- \pi^+)$ from Eq.
(2.21) by combining both the $D$ and $D^+_s$ yields in tagged
$\overline B$ decays.

\newpage

\begin{center}
\bf{Comparison of the Theory of Inclusive $B$ Decays with Experimental Results}
\end{center}

The direct $B(b\rightarrow c\bar cs^\prime )$ measurement is now known to be
compatible with theoretical predictions ($0.24 \pm 0.051$ in on-shell scheme and
$0.30 \pm 0.044$ in $\overline{MS}$ scheme) \cite{bagan,voloshin}, which have
larger uncertainties. Furthermore, we are finally in a situation to confront
the theoretical calculation of 4.0 $\pm$ 0.4 \cite{baganrud} for
\begin{equation}
r_{ud} \equiv \frac{\Gamma (b\rightarrow c\bar ud^\prime )}{\Gamma
(b\rightarrow
ce\bar\nu )} \;,
\end{equation}
which served as a necessary input for a number of our predictions \cite{bdy}.
With accurate
measurements of $B(b\rightarrow c\bar cs^\prime )$ and $B(\overline
B\rightarrow X
e\bar\nu )$, $r_{ud}$ is determined by \cite{fwd,bsbsbar,bdy},
\begin{equation}
r_{ud}\bigg|_{exp} = \frac{1-B(b\rightarrow c\bar cs^\prime
)}{B(\overline
B\rightarrow X e\bar\nu )} -(2 +r_\tau +r_{\not c})\;.
\end{equation}
We use \cite{alephtau,flnntau}
\begin{equation}
r_\tau \equiv \frac{\Gamma (b\rightarrow c\tau\bar\nu )}{\Gamma
(b\rightarrow
ce\bar\nu )} = 0.25 \;,
\end{equation}
\begin{equation}
B(\overline B\rightarrow X \ell\bar\nu )=(10.49 \pm 0.46)\% \;
(CLEO\;\; II),
\end{equation}
and we have chosen \cite{bdy}
\begin{equation}
r_{\not c} \equiv \frac{\Gamma (b\rightarrow \;{\rm no}\;{\rm
charm})}{\Gamma
(b\rightarrow ce\bar\nu )} = 0.25 \pm 0.10 \;  ,
\end{equation}
to obtain
\begin{equation}
r_{ud}\bigg|_{exp} = 4.86 \pm 0.47 \;.
\end{equation}
If the discrepancy between the measured and theoretically predicted values of
$r_{ud}$ persist, it will indicate
either large higher order QCD corrections~\cite{voloshinqcd} or
important nonperturbative effects
at ${\cal O} (1/m^3_b )$~\cite{neuberts,bbd} or both.

We favor the explanation due to non-perturbative effects at ${\cal O} (1/m^3_b )$. The reason goes as follows. There are no ${\cal O} (1/m^3_b )$ corrections of the Pauli-interference and Weak-annihilation type for $\overline B$ decays governed by $b\rightarrow c\bar cs$ transitions. Thus,
comparing theory and experiment for $B(b\rightarrow c\bar cs^\prime)$
indicates whether higher order QCD corrections are important for $b\rightarrow
c\bar cs^\prime$ transitions, under the assumption of local duality. Because theory
and experiment are compatible for $B(b\rightarrow c\bar cs^\prime )$ (with theory
tending to give slightly larger values), we speculate that ${\cal O} (1/m^3_b )$
effects rather than higher order
QCD corrections enhance the $b\rightarrow c\bar ud^\prime$ rate.

We conclude that there is no problem between the precisely
measured $B(\overline B\rightarrow X\ell\bar\nu )$ and the state of the art
theoretical investigations~\cite{neuberts}. In fact, CLEO's recent measurements provide theorists
with several consistency checks. A successful theoretical calculation must not only
agree with the precisely measured $B(\overline B\rightarrow X\ell\bar\nu )$, but
with the following accurately-known quantities as well:
\begin{equation}
B(b\rightarrow c\bar cs^\prime ) = 0.227 \pm 0.035, \;{\rm and}
\end{equation}
\begin{equation}
r_{ud} = 4.86 \pm 0.47 \;.
\end{equation}

\section{The Low $R_c$ Measurement}

Whereas theory predicts
\begin{equation}
R_c \equiv \frac{\Gamma (Z^0 \rightarrow c\bar c)}{\Gamma (Z^0
\rightarrow
\;{\rm hadrons})} = 0.172 \;,
\end{equation}
LEP/SLC have measured it 2 $\sigma$ below this value at \cite{graziani}
\begin{equation}
R_c |_{exp} = 0.1598 \pm 0.0069 \;.
\end{equation}

One must, however, distinguish among the various $R_c$ measurements.
The measurements which fully reconstruct primary $D^{*+}$'s are
inversely proportional to $B(D^0 \rightarrow K^- \pi^+)$

\begin{equation}
R_c (DELPHI \;D^*) = 0.148 \pm 0.007 \pm 0.011 \;\cite{delphirc},
\end{equation}
\begin{equation}
R_c (OPAL\;D^*) = 0.1555 \pm 0.0196 \;\cite{behnke},
\end{equation}
with a world average of
\begin{equation}
R_c (D^*)= 0.150 \pm 0.011 \;.
\end{equation}

Unfortunately OPAL and DELPHI have not explicitly presented the
uncertainty due to
$B(D^0\rightarrow K^- \pi^+)$ in their $R_c$ measurements which would
allow $B(D^0
\rightarrow K^-\pi^+)$ to be determined straightforwardly.  In the
absence of such information, we will be conservative and retain the
full uncertainty in $R_c$ to obtain
\begin{equation}
17.2  = (15.0 \pm 1.1)
\left[\frac{3.84\%}{B(D^0
\rightarrow K^- \pi^+)}\right]\;.
\end{equation}
This yields
\begin{equation}
B(D^0 \rightarrow K^-\pi^+) =  (3.35 \pm 0.25)\% \; ,
\end{equation}
which is compatible with our extracted value of
$B(D^0 \rightarrow K^-\pi^+)$ from $n_c$ (see Eq.~(2.26)).

DELPHI also measured $R_c$ via an inclusive double tag method, where only
the daughter pions of the energetic $D^{*\pm}$ have been reconstructed.
This method does not involve $B(D^0 \rightarrow K^- \pi^+)$ and
the result, albeit with large uncertainty,
\cite{graziani}
\begin{equation}
R_c (\pi^+\pi^-) = 0.171^{\textstyle{+0.014}}_{\textstyle{-0.012}}
\pm 0.015
\end{equation}
agrees with theory without modification.

There is also a lepton method for measuring $R_c$, but it has very large
systematic uncertainties, and
there are $R_c$ measurements from both OPAL and DELPHI from direct
charm counting \cite{graziani,opalrc}. The extraction of
$B(D^0 \rightarrow K^- \pi^+)$  from the latter
is less direct than from the fully reconstructed $D^{*+}$ data
samples, because additional charmed hadrons are involved.
We illustrate what is involved with
the data sample published recently by OPAL~\cite{opalrc},

\begin{equation}
R_c ({\rm charm}\;{\rm counting}) =  0.167 \pm 0.011\;(stat) \pm
0.011\;(sys) \pm 0.005\;(br)\;.
\end{equation}
OPAL measured
\begin{eqnarray}
R_c \;f(c\rightarrow D^0)\;B(D^0 \rightarrow K^- \pi^+) & = & (0.389
\pm 0.037) \%
\;,\nonumber \\
R_c \;f(c\rightarrow D^+) \;B(D^+\rightarrow K^- \pi^+\pi^+) & = &
(0.358 \pm 0.055) \%
\;, \nonumber \\
R_c \;f(c\rightarrow D^+_s) \;B(D^+_s \rightarrow \phi\pi^+) & = &
(0.056 \pm 0.017) \%
\;, \nonumber \\
R_c \;f(c\rightarrow \Lambda^+_c ) \;B(\Lambda^+_c \rightarrow pK^-
\pi^+ ) & = &
(0.041 \pm 0.020) \% \;.
\end{eqnarray}
It then summed these fractions using the reference branching fractions
from the preliminary 1996 PDG:
\begin{eqnarray}
B(D^0 \rightarrow K^- \pi^+) & = & (3.84 \pm 0.13)\% \;, \nonumber \\
B(D^+\rightarrow K^- \pi^+ \pi^+ ) & = & (9.1 \pm 0.6)\% \;,
\nonumber \\
B(D^+_s \rightarrow \phi\pi^+) & = & (3.5 \pm 0.4)\% \;, \nonumber \\
B(\Lambda_c \rightarrow pK^- \pi^+) & = & (4.4 \pm 0.6)\% \;.
\end{eqnarray}
OPAL assumed that the undetected primary $\Xi_c$ and $\Omega_c$
production is $(15 \pm 5) \%$ of the primary $\Lambda_c$ production,
and thus
obtained Eq.~(3.9).

We want to modify this treatment in several respects. First,
we wish to
solve for $B(D^0 \rightarrow K^- \pi^+)$ assuming the standard model
value for
$R_c = 0.172$.
Second, as Section VII explains, a more satisfactory
estimate for
$B(\Lambda_c\rightarrow pK^- \pi^+)$ is $(6.0 \pm 1.5)\%$, rather
than $(4.4 \pm 0.6)\%$.
This causes the primary production fraction of $\Lambda_c$ to decrease.
 We then correlate the inclusive primary production
fraction of $baryon_c$ to that of $\Lambda_c$ via
\begin{equation}
f(c\rightarrow baryon_c ) =  \;f(c\rightarrow \Lambda_c )\;/ \;
(1-p)^2\;,
\end{equation}
where $p$ models the production fraction of $s\bar s$ fragmentation
relative to $f\bar f$ from the vacuum, where $f=u, d$ or $s$
\cite{pvalue}.
A value of $B(D^0 \rightarrow K^- \pi^+)$ is thus obtained via
\begin{equation}
B(D^0 \rightarrow K^- \pi^+) = \frac{(0.389 \pm 0.037 ) + \frac{0.358
\pm
0.055}{r_+}}{R_c - \frac{(0.056 \pm 0.017)\%}{B(D_s
\rightarrow \phi\pi^+)} - \frac{(0.041 \pm 0.020)\%}{(1-p)^2
B(\Lambda_c\rightarrow
pK^- \pi^+ )}}\;\; \% .
\end{equation}
Inserting,
\begin{equation}
r_+ = 2.35 \pm 0.23, R_c = 17.2\%, B(D_s \rightarrow \phi\pi^+) =
(3.5 \pm 0.4)\%
\;,
\end{equation}

\begin{equation}
{\rm
and}\;B(\Lambda_c \rightarrow pK^- \pi^+ ) = (6.0 \pm 1.5 )\%,
\end{equation}
we obtain
\begin{equation}
B(D^0 \rightarrow K^- \pi^+) = (3.67 \pm 0.42)\% \;.
\end{equation}
This is again compatible with our other $B(D^0
\rightarrow K^- \pi^+)$ determinations.
We have thus shown that a reduction in the value of
$B(D^0 \rightarrow K^- \pi^+ )$  eliminates
the discrepancy between theory and experiment regarding $R_c$.

\section{Anatomy of $\overline B\rightarrow D\overline D \;\overline
KX$ Processes}

Since CLEO II has proven the prediction \cite{bdy} that a sizable
fraction of all
$B$ decays are governed by $\overline B\rightarrow D\overline D
\;\overline KX$, it is imperative to classify these processes.
This section serves then the dual purpose of
aiding experimentalists in finding and classifying the $\overline
B\rightarrow D\overline D \;\overline KX$ processes and discusses how
to properly take them into account in other $B$ decay analyses, such as
semileptonic $B$ decays, where they represent a background.

Current [semileptonic] analyses have not accounted for
$\overline B\rightarrow D\overline D \;\overline K X$ backgrounds,
and therefore need to be modified, as discussed in the next section.
This section focusses on the more global properties of the $\overline
B\rightarrow D\overline D \;\overline KX$ transitions.

Isospin symmetry is a powerful tool for these processes
because the underlying $b\rightarrow c\bar cs$ quark transition has $I=0$.
A sizable [probably majority] fraction of all $\overline
B\rightarrow D\overline D \;\overline K X$ decays will be of the exclusive
form $\overline B\rightarrow D^{(*)} \overline D^{(*)} \overline K$
\cite{nokstar},
\begin{eqnarray}
B^- \rightarrow && D^{(*)+} \;D^{(*)-} \;K^- \;, \nonumber \\
&& D^{(*)0} \;\overline D^{(*)0} \;K^- \;,\nonumber \\
&& D^{(*)0} \;D^{(*)-} \;\overline K^0 \;,
\end{eqnarray}
\begin{eqnarray}
\overline B_d \rightarrow && D^{(*)+} \;D^{(*)-} \;\overline K^0
\;,\nonumber \\
&& D^{(*)+} \;\overline D^{(*)0} \;K^- \;,\nonumber \\
&& D^{(*)0} \;\overline D^{(*)0} \;\overline K^0 \;.
\end{eqnarray}
Isospin symmetry alone demands that \cite{lipkinsanda,june95}
\begin{equation}
\Gamma (B^- \rightarrow D^{(*)+} \;D^{(*)-} \;K^-) = \Gamma
(\overline B_d
\rightarrow D^{(*)0} \;\overline D^{(*)0} \;\overline K^0) \;,
\end{equation}
\begin{equation}
\Gamma(B^- \rightarrow D^{(*)0} \;\overline D^{(*)0} \;K^-) = \Gamma
(\overline
B_d \rightarrow D^{(*)+} \;D^{(*)-} \;\overline K^0) \;,
\end{equation}
\begin{equation}
\Gamma (B^- \rightarrow D^{(*)0} \;D^{(*)-} \;\overline K^0) = \Gamma
(\overline B_d \rightarrow D^{(*)+} \;\overline D^{(*)0} \;K^- ) \;.
\end{equation}
Color transparency arguments
\cite{bjorkencolortransparency} predict that the isospin of
$\overline D^{(*)}
\overline K$ is essentially zero \cite{june95}
\begin{equation}
B(b \to c + (\overline D^{(*)}\overline K)_{I = 0}) \;\; \gg
B(b \to c + (\overline D^{(*)}\overline K)_{I = 1})
\end{equation}
This could also be demonstrated by observing
that the virtual $W\rightarrow \bar cs$ decay gives rise to an isospin and color
singlet. The isospin singlet, $I_{\bar cs}=0$,
hadronizes independent of what the rest of the system does under
the factorization assumption \cite{bsw}, thus corroborating Eq.~(4.6).
It then follows that \cite{june95}
\begin{equation}
\Gamma (B^- \rightarrow D^{(*)+} \;D^{(*)-} \; K^-) = \Gamma
(\overline B_d \rightarrow D^{(*)0} \;\overline D^{(*)0} \;\overline K^0) =0 \;,
\end{equation}
and for the remaining processes,
\begin{eqnarray}
\Gamma (B^- \rightarrow D^{(*)0} \;\overline D^{(*)0} \; K^-) & = &
\Gamma (B^- \rightarrow
D^{(*)0} \;D^{(*)-} \; \overline K^0) = \nonumber \\
=\Gamma (\overline B_d \rightarrow D^{(*)+} \;D^{(*)-} \; \overline
K^0) & = &
\Gamma (\overline B_d \rightarrow D^{(*)+} \;\overline D^{(*)0} \;
K^-)\;.
\end{eqnarray}
Of the 24 potentially different rates represented in Eqs.
(4.1)-(4.2), we have thus reduced the
problem to 4 (as yet) unrelated, reduced matrix elements,
\begin{equation}
\overline B \rightarrow D^{(*)} \;\overline D^{(*)} \; \overline K \;.
\end{equation}
To go further, we complete the implications of the factorization
assumption, wherein the
[relative] $D$ and $D^*$
production is described by the $\overline B\rightarrow D^{(*)}$ form
factors \cite{hqet,bcincl}.
To probe the hadronization of $\bar cs$ into $\overline D^{(*)}
\overline K$, one may contemplate the matrix element
\begin{equation}
<\overline D^{(*)} \overline K |\bar s\gamma_\mu (1-\gamma_5 ) c|0> \;.
\end{equation}
This presents a formidable, but $academic$, theoretical problem. The dominant
hadronization processes will be $\bar cs \rightarrow D^{(*)-}_s$ and
$\bar cs \rightarrow D^{r-}_s \rightarrow \overline D^{(*)} \overline KX$, where
$D^{r-}_s$ denotes all $\bar
cs$ resonances beyond the $D^-_s$ and $D^{*-}_s$. The following
matrix elements are thus of great importance
\begin{equation}
<D^{r-}_s |\bar s\gamma_\mu (1-\gamma_5 )c|0> \;,
\end{equation}
and are being analyzed at present \cite{veselid}.
Because of the V-A nature of the current, the final $D_s$ resonances cannot
have spin 2 or higher. Spinless $p$ wave resonances are suppressed, as can be
seen by taking the limit $m_s
\rightarrow m_c$ in
which case the matrix element [Eq.~(4.11)] vanishes.
We expect the radially excited s wave $D_s$ resonances, $0^-$ and
$1^-$, to be significant contributors, because their decay constants
are found to be very large in preliminary lattice studies~\cite{eichten}.

Whereas the $b \to c + \overline D^{(*)} \overline K$ processes were
the highlight of this section, in the next section they will be viewed
as a background in semileptonic $B$ decays.

\section{Semileptonic $B$ Decays}

Semileptonic decays are one of the most studied aspects of $B$ hadrons
~\cite{burchatr}. These must be reevaluated since roughly 10\%
of all B's decay via $\overline B\rightarrow D\overline D\;\overline K X$
\cite{bdy} which introduce a $\overline B \to
\overline D \to \ell^-$ background that has not previously been considered.
We will discuss inclusive decays first, followed by a detailed
accounting of their various exclusive components.

\subsection{Inclusive Semileptonic $B$ Decays}

Inclusive semileptonic measurements include the single
lepton analyses and the so-called ``model independent" dilepton analyses
\cite{burchatr,browderh95,wang}.
The lepton spectrum in these analyses is made up of the following
components: primary leptons from $(b\rightarrow c\ell\nu )$, secondary [or
cascade] leptons from $(b\rightarrow cX, c\rightarrow s\ell\nu ),$ and
leptons from primary charm decays $(c\rightarrow s\ell\nu )$.
Because $|V_{ub} / V_{cb}| \approx 0.1$, the $b \to u \ell \nu$
transition is highly suppressed.

One of the well known features of the V-A interactions is that,
in the respective restframes of the decaying heavy flavors, the
charged lepton spectrum for $b\rightarrow q\ell^- \bar\nu$ transitions is
hard,  whereas that of the $c\rightarrow q^\prime \ell^+ \nu$ transitions,
is soft. Together with the fact that $M_b > M_c$, it follows that
the primary leptons $(b\rightarrow q \ell^- )$ will be harder
[in momentum $p$ at $\Upsilon
(4S)$ factories, in $p$ as well as $p_{T,rel}$ at higher energy colliders]
than the secondary leptons $(b\rightarrow c\rightarrow \ell^+ ,
b\rightarrow \bar c \rightarrow \ell^- )$. In the laboratory frame
the secondary charm has a boost that must be taken into
account and the $B$ mesons are not strictly at rest
in the center of mass frame of an $\Upsilon (4S)$.

Figure 2 of Ref.~\cite{cleoslincl} shows the inclusive lepton spectra at the $\Upsilon (4S)$.
 That
Figure demonstrates that essentially only primary leptons satisfy
the $p > 1.5$ GeV/c cut, whereas both secondary and primary leptons contribute
at lower momenta.

\subsubsection{Single Inclusive Lepton Analysis at the $\Upsilon (4S)$}

In many of the single lepton analyses, the shape of the cascade lepton
momentum spectrum is obtained by
convoluting the measured $\overline B\rightarrow D X$ momentum
distribution with that of the $D \rightarrow \ell X$.
Secondary leptons from cascading $D_s , \Lambda_c$, and $J/\psi$
decays ($\overline B\rightarrow (\Lambda_c , D^-_s , J/\psi)
\rightarrow \ell$)  are treated as background and are subtracted.
Since no charge correlations are performed in the analysis, it is
immaterial whether the regular $D$ (i.e., $D^0$ and $D^+$) charmed hadrons
are created through $b\rightarrow c$ and/or $b\rightarrow \bar c$ transitions.
The normalizations of the various lepton components are extracted from a
fit to the inclusive lepton data sample. Ultimately, significant model
dependence persists in the determination of the inclusive
primary $B(\overline B\rightarrow X\ell^- \bar\nu )$ as Table IV demonstrates.

It is therefore important to note that ARGUS invented a ``model independent"
dilepton method, which extracts the inclusive yields of the primary and
secondary components separately.  CLEO improved upon this method by
introducing the more optimal Wang diagonal cut \cite{wang,cleoslincl}.
ALEPH tailored the dilepton analysis to the $Z^0$ environment.
However, as mentioned above, the dilepton analyses
are flawed in two respects. First, the
background from $\overline B\rightarrow \overline D\rightarrow \ell^-$
transitions was not taken into account. Second, the removal
of $B-\overline B$ mixing effects is
more subtle than has been assumed. The next subsection describes how
to incorporate these two effects and attempts to quantify the
subsequent modifications.

\subsubsection{Model Independent Dilepton Analysis}

We now take issue with the so-called ``model independent"
dilepton method invented by ARGUS
\cite{argusdilepton}, and improved by CLEO \cite{wang,cleoslincl}.
First, the dilepton analysis has to be corrected for the neglected
$\overline B\rightarrow \overline D \rightarrow e^-$ background.
The current procedure takes the lepton from the decay of the
wrong-charmed hadron $(\overline B\rightarrow \overline D \rightarrow e^- )$
to be primary, thereby deceptively increasing the primary component.
Second, the current modelling of $B^0-\overline B^0$ mixing effects is
flawed in that it implicitly assumes that
$$B(\overline B_d \rightarrow D \rightarrow e^+ X)=B(B^- \rightarrow
D\rightarrow e^+ X).$$
The correct removal of $B^0 -\overline B^0 $ mixing effects is
discussed in Ref.~\cite{distinguish}.
It is straightforward once we recognize that
$$B(\overline B_d \rightarrow D\rightarrow e^+ X) > B(B^- \rightarrow
D\rightarrow e^+ X).$$
Before we begin a more detailed discussion of these modifications, we
will briefly review the current dilepton analyses.

Strict cuts on the first lepton guarantee it to be primary
$(b\rightarrow  \ell^-)$. This is the lepton used in the ``tag" of one $B$
in the event. No momentum restrictions are placed on the second lepton, which
in the case  of ARGUS and CLEO is an $e^\pm$. Angular correlations, however,
are used to ensure that the second lepton comes from the other
$B$ in the event. The unlike-sign and
like-sign lepton momentum spectra are expressed in terms of the primary
($B(b)$), and cascade ($B(c)$), branching fractions
\cite{argusdilepton,wang,cleoslincl}.
\begin{equation}
\frac{dN_{\pm\mp}}{dp} \sim \epsilon (p) \left(\frac{dB(b)}{dp}
\;(1-\chi
)+\frac{dB(c)}{dp} \chi\right) \;,
\end{equation}
\begin{equation}
\frac{dN_{\pm\pm}}{dp} \sim \frac{dB(b)}{dp} \chi +\frac{dB(c)}{dp}
\;(1-\chi )\;.
\end{equation}
Here $\epsilon (p)$ is the momentum dependent efficiency of a cut that removes
unlike-sign dileptons originating from a single $B$ decay and $\chi$
parameterizes $B^0 -\overline B^0$ mixing.
The primary and secondary electron spectra were obtained by solving
these two equations. It is claimed that there is no model dependence for
the measured momentum spectrum $\frac{dB(B\rightarrow Xe\nu )}{dp}$,
where $P > 0.6$ GeV/c
\cite{cleoslincl}. Refs.~\cite{wang,cleoslincl} have subtracted
backgrounds coming from inclusive charmed baryon and $D_s$ production in B
decays.

\begin{center}
\bf{a. The $\overline B\rightarrow \overline D  \to \ell^-$
Background}
\end{center}

The background from $\overline B\rightarrow \overline D  \to \ell^-$
transitions however, was not taken into account.  To clarify this criticism
note that if there were no mixing ($\chi =0$),  the
$\overline B\rightarrow \overline D  \to  \ell^-$
transitions would feed into the unlike-sign dilepton data sample.
The ``model independent" analysis is therefore more complicated than
currently believed, but nevertheless possible once one differentiates
between the conventional $b\rightarrow c\rightarrow e^+$, and the
additional $b\rightarrow \bar c\rightarrow e^-$, secondary lepton
sources.

To accurately account for the background due to $\overline
B\rightarrow \overline D  \to \ell^-$,
one has to study the $D-\ell^\pm$ and $D^* -\ell^\pm$ correlations,
where the $D^{(*)}$ and lepton have different $B$ parents. In such a study,
one must

\begin{enumerate}
\begin{enumerate}
\item remove $B_d -\overline B_d$ mixing effects, and
\item determine the probability that a wrong-sign charm
$(\overline D)$ is
seen as a $\overline D^0$ as opposed to a $D^-$ [since their semileptonic
branching fractions differ].
\end{enumerate}
\end{enumerate}

Let us explain what is involved in our case. Denote by $D^{(*)}$ the sum of the charged $D^{(*)+}$
and neutral $D^{(*)0}$. The inclusive wrong-charm rate is predicted
to satisfy,
\begin{equation}
\Gamma (B^- \rightarrow \overline D^{(*)} X) =\Gamma (\overline B_d
\rightarrow
\overline D^{(*)} X) \;,
\end{equation}
whereas the right-charm rate satisfies
\begin{equation}
\Gamma(B^-\rightarrow D^{(*)} X)\approx \Gamma (\overline B_d
\rightarrow
D^{(*)} X) \;.
\end{equation}
While the color allowed and color suppressed $B$ decay amplitudes
interfere for the $B^-$, they do not for the neutral $\overline B_d$.
Given the inclusive nature of the processes under consideration, it
is expected that the above approximation will be still valid. Since the
charged and neutral $B$  lifetimes are approximately
equal \cite{bigistone,schune},
\begin{equation}
B(B^- \rightarrow \overline D^{(*)} X) \approx B(\overline B_d
\rightarrow
\overline D^{(*)} X) \;,
\end{equation}
\begin{equation}
B(B^- \rightarrow  D^{(*)} X) \approx B(\overline B_d \rightarrow
 D^{(*)} X) \;,
\end{equation}
the $B_d -\overline B_d$ mixing removal is now straightforward from
 measurements of the $\ell^\pm -D$ and $\ell^\pm -\overline D^{(*)}$
correlations, separately \cite{distinguish} (where the hard primary lepton comes
from one $B$ and the charmed hadron from the other $B$ in the process).

With mixing successfully removed, one has the separate inclusive BR's
into wrong-charmed $\overline D$ and $\overline D^*$,
for an unmixed $\overline B$. Isospin symmetry tells us that
the wrong-charm $\overline D^*$'s are seen in equal fractions as
$D^{*-}$ and $\overline D^{*o}$, which is also true for
the wrong-charmed $\overline D$'s that do not originate from $\overline D^*$'s !

Note that the procedure just described ignores (a) kinematic threshold
effects where $\bar cs \rightarrow D^{**-}_s$ in the neighborhood of the
$\overline D^{(*)}\overline K$ mass, and (b) the Cabibbo suppressed transition
$W\rightarrow \bar cd\rightarrow D^{(*,**)-}$. These small effects can be
incorporated if desired.
The relative fractions of the wrong-charmed mesons hadronizing into
$D^-$ and $\overline D^0$ can therefore be experimentally determined to
allow the accurate modelling of the
$\overline B \to \overline D \to \ell^-$ background.

If all other backgrounds were modelled correctly, the subtraction of
the $\overline B\rightarrow \overline D \rightarrow \ell^-$ background
would decrease the semileptonic BR of $B$ mesons since, as discussed above,
the various analyses have taken the lepton from the decay of the wrong-
charmed hadron $(\overline B\rightarrow \overline D \rightarrow\ell^- )$ to
be primary which incorrectly increased the primary component.
We suspect however that CLEO has oversubtracted some of the
other $b\rightarrow \bar cs \rightarrow e^-$ backgrounds.  It may
therefore be worthwhile to quantify our expectations.

An elaborate account of the ``model independent" inclusive semileptonic
$B(\overline B\rightarrow X\ell\bar\nu )$ measurement can be found in
Roy Wang's thesis \cite{wang} and in Ref.~\cite{cleoslincl}.
Since the predominant representative of the $\bar cs$ background (in
$b\rightarrow c\bar cs$ transitions) is $D^-_s$, we needed to
understand it and therefore
obtained the following central value parameterization \cite{wangthank}:
\begin{equation}
\int^{2.6 \;{\rm GeV/c}}_{0.6 \;{\rm GeV/c}}dp_e \;\;\frac{dB}{dp_e}
(\overline
B\rightarrow Xe\bar\nu ) = \left[9.85 +0.434 \left(1-w\right)\right]
\% \;.
\end{equation}
Here $w$ denotes the actual $b\rightarrow \bar cs\rightarrow e^-$
background in units of the original $D_s$ background used by Wang in his thesis,
\begin{equation}
w (s)\equiv \frac{s\cdot B(b\rightarrow \bar cs\rightarrow
e^-)|_{\overline
B\rightarrow D\overline D\;\overline KX} + B(b\rightarrow \bar
cs\rightarrow e^-
)|_{\overline B\rightarrow D^-_s X}}{B(b\rightarrow \bar
cs\rightarrow e^-
)|_{\overline B\rightarrow D^-_s X \;{\rm [Wang]}}} \;.
\end{equation}
The quantity $s = 0.2 \pm 0.1$ is a correction factor, which takes
into account the fact
that the diagonal cut invented by Wang \cite{wang} suppresses the
$B(b\rightarrow \bar cs\rightarrow e^-)|_{\overline B \rightarrow D
\overline D\;\overline K X}$ background more than that of the
$B(b\rightarrow \bar cs\rightarrow e^-)|_{\overline B \rightarrow
D^-_s X}$, on account of its softer lepton spectrum.
In general~\cite{neglectccdbkgd}
\begin{eqnarray}
&& B(b\rightarrow \bar cs \rightarrow  \ell^- )|_{\overline
B\rightarrow
D\overline D \;\overline KX} = \nonumber \\
& = & \frac{1}{2} \;\frac{B(\overline B\rightarrow \overline
DX)}{(1+r)}
\;B(\overline D^0 \rightarrow \ell^- X) \bigg\{ 1+r+B\left(D^{*-}
\rightarrow \overline D^0 \pi^- \right) + \nonumber \\
& + & \frac{B(D^- \rightarrow \ell^- X)}{B(\overline D^0 \rightarrow
\ell^-
X)} \;\left[r+B\left(D^{*-} \rightarrow D^- X^0 \right)\right]\bigg\}
\approx   \nonumber \\
& \approx & \frac{1}{2} \;\frac{B(\overline B\rightarrow \overline
DX)}{(1+r)}
\;B(\overline D^0 \rightarrow \ell^- X) \bigg\{
1+r+B(D^{*-}\rightarrow
\overline D^0 \pi^- ) + \nonumber \\
& + & \frac{\tau (D^+)}{\tau (D^0)} \;\left[r+B\left(D^{*-}
\rightarrow D^-
X^0 \right)\right]\bigg\} \;.
\end{eqnarray}
The last approximation is excellent and assumes the same inclusive
semileptonic rates for $D^-$ and $\overline D^0$.  The observable
$r$ denotes
\begin{equation}
r\equiv \frac{B(\overline B\rightarrow X+ \; \overline D_{dir})}
{B(\overline B\rightarrow X+ \overline D^*)} \;,
\end{equation}
where $\overline D_{dir}$ denotes wrong-charm $\overline D$ without
$\overline D^{*}$
parentage. As demonstrated above, $r$ can be experimentally
determined from $\ell^\pm - D$ and
$\ell^\pm - D^*$ correlations.  Whereas the $D_s$ background in
Wang's thesis was taken to be
\cite{wang,wangthank},
\begin{equation}
B(b\rightarrow \bar cs \rightarrow e^- )|_{\overline B\rightarrow
D^-_s X
{\rm [Wang]}} =
B(\overline B\rightarrow D^-_s X) B(D^-_s \rightarrow  Xe^- \bar\nu )
= 0.1181 \times 0.0793 = 9.36 \times 10^{-3} \;,
\end{equation}
we use a smaller inclusive semielectronic $D_s$ BR,
\begin{eqnarray}
B(D^-_s \rightarrow Xe^- \bar\nu ) & = & \frac{\Gamma (D^-_s
\rightarrow X e^-
\bar\nu )}{\Gamma (\overline D^0 \rightarrow Xe^- \bar\nu )}
\;\;\frac{\tau
(D_s)}{\tau (D^0)} \;\;B(\overline D^0 \rightarrow Xe^- \bar\nu )
\approx
\nonumber \\
& \approx & B(\overline D^0 \rightarrow Xe^- \bar\nu )=(6.64 \pm 0.18
\pm 0.29 )\%\; .
\end{eqnarray}
The reason is that while the lifetime ratio is measured as
\cite{pdg},
\begin{equation}
\frac{\tau (D_s)}{\tau (D^0 )} = 1.12 \pm 0.05
\end{equation}
ISGW2 \cite{isgw2} predicts a substantial decrease in the inclusive
semielectronic $D_s$ rate versus that of the $D^0$, because of the
restricted phase space of $D_s \rightarrow \eta^\prime e^- \bar\nu$.

The portion of the inclusive $D_s$ yield in $B$ decays that contributes to
the background is \cite{chomenary}
\begin{equation}
B(\overline B\rightarrow D^-_s X) = 0.100 \pm 0.017 \;,
\end{equation}
so that we obtain
\begin{equation}
B(b\rightarrow \bar cs\rightarrow e^- )|_{\overline B\rightarrow
D^-_s X} =
B(\overline B\rightarrow D^-_s X) B(D^-_s \rightarrow  Xe^- \bar\nu )
\approx
6.64 \times 10^{-3} \;.
\end{equation}

Table V estimates
$B(b\rightarrow \bar cs \rightarrow e^- )|_{\overline B\rightarrow
D\overline D\;\overline KX} \;$ and $\;w ,$
as a function of $r$.
It uses the recent CLEO measurement \cite{freyberger},
$$B(\overline D^0 \rightarrow Xe^- \bar\nu )= (6.64 \pm 0.18 \pm
0.29) \;\%\;,$$
and takes  the $D^+/D^0$ lifetime ratio
and  $B(D^{*-} \rightarrow \overline D^0 \pi^- )$ from the 1994
particle data group
\cite{pdg}. Our guess for the ``actual" $b\rightarrow \bar
cs\rightarrow e^-$ background is
roughly as large as that of the original $D^-_s$ background employed in the
published CLEO II analysis
\cite{cleoslincl} and in Wang's thesis \cite{wang}. It is relatively
insensitive to the precise value chosen for $r$, as seen in Table V.
For $ w= 0.9$, the primary lepton BR between $0.6 \leq p_e \leq 2.6$
GeV/c is [see Eq.~(5.7)],
\begin{equation}
B(\overline B\rightarrow Xe^- \bar\nu ,p_e \geq 0.6\; {\rm GeV/c}) =
9.89\%
\;. \end{equation}
The undetected primary fraction is estimated to be~\cite{cleoslincl},
\begin{equation}
\frac{B(\overline B\rightarrow Xe^- \bar\nu , \;p_e < 0.6\; {\rm
GeV/c})}{B(\overline B\rightarrow Xe^- \bar\nu )} = (6.1 \pm 0.5)\%\;.
\end{equation}
We therefore interpret the published CLEO II data
\cite{cleoslincl,wang} to mean
\begin{equation}
B(\overline B\rightarrow Xe^- \bar\nu )=(10.5 \pm 0.5)\%\;,
\end{equation}
which is in excellent agreement with the published CLEO II result
\begin{equation}
B(\overline B\rightarrow Xe^- \bar\nu )= (10.49 \pm 0.46)\%\;.
\end{equation}

We found that the impact of the $\overline B\rightarrow \overline D
\rightarrow e^-$ background is much reduced on account of its
soft momentum spectrum, which causes it to be efficiently removed by the Wang
cut~\cite{wang}.  Because CLEO has probably oversubtracted the
$\overline B\rightarrow D^-_s \rightarrow e^-$ background, not much changed in
overall normalization when we added the $\overline B\rightarrow \overline D
\rightarrow e^-$ background to our $\overline B\rightarrow D^-_s
\rightarrow e^-$ estimate.  However, we predict the $b\rightarrow
\bar cs \rightarrow e^-$ background
to be softer than that which they used, resulting in a stiffer
primary lepton spectrum than currently measured by
CLEO~\cite{wang,cleoslincl}.

Clearly, other backgrounds must be modified.  For instance, the ``$e$
from same $B$ background" listed in Table
4.5 of Wang's thesis is predicted to increase because of an expected
increase in the exclusive $B(\overline B\rightarrow D^{(*)} \ell \bar\nu$)
which in turn is due to an expected decrease in $B(D^0 \rightarrow K^- \pi^+ )$.

We encourage CLEO to carry out the necessary modifications.
We have seen that the $\overline B\rightarrow \overline D \rightarrow e^-$
background is much reduced for the present analyses at
symmetric $\Upsilon (4S)$ factories, primarily because of a
fortuitous momentum cut $p_e >0.6$ GeV/c. At asymmetric
$\Upsilon (4S)$ factories and for $Z^0$ factories similar reductions do
not occur and so this background will contribute
significantly and must be carefully subtracted. It is
partially responsible for the apparent larger $B(b\rightarrow X\ell
\bar\nu )$ at $Z^0$ factories compared to $\Upsilon (4S)$ machines,
see Section V.A.3  below.

Before moving on to $Z^0$ factories, we will next discuss
possible flaws in the treatment $B^0-\overline B^0$ mixing effects
by ARGUS and CLEO.

\begin{center}
\bf{b. Removal of $B^0 -\overline B^0$ Mixing Effects}
\end{center}

We believe that the removal of $B^0 -\overline B^0$ mixing effects is not
correctly performed in experimental analyses. CLEO \cite{crawford}, for
instance, has removed mixing from their observed  $\Lambda
-\ell^\pm$ data sample by adding and subtracting a constant.
ARGUS \cite{argusbaryon} has implicitly assumed equal inclusive baryon
production fractions from $B_d$ and $B^+$ decays separately.
The same implicit assumption was made by CLEO when $B(\overline
B\rightarrow
\overline\Lambda_c X)/B(\overline B\rightarrow \Lambda_c X) = 0.20
\pm 0.14$ was determined from $\ell^\pm \Lambda_c$ correlations
\cite{glasgowbaryon}.
The quoted error does not include the following systematic
uncertainty. Suppose that only charged $B$'s produce
$\stackrel{(-)}{\Lambda_c}$ baryons. Then the $\ell^\pm \Lambda_c$
correlations should clearly not be corrected for $B^0 -\overline B^0$ mixing
effects. On the other hand, if only neutral $B$'s produce
$\stackrel{(-)}{\Lambda_c}$,
then $B^0 -\overline B^0$ mixing effects are maximal and must be
removed.

Ref.~\cite{distinguish} discusses how to properly take into account
otherwise confusing $B^0 -\overline B^0$ mixing effects. We will briefly
summarize the procedure here.
The charged and neutral $B$ meson lifetimes and production rates are
currently found to be approximately equivalent and will be assumed
to be identical. (It is a straightforward exercise to incorporate
inequalities if such are observed.)
Suppose that one wishes to determine $B(B\rightarrow TX)$ and
$B(\overline B\rightarrow TX)$, where $T$ denotes any flavor specific
partially  reconstructed final state.  $N_{\ell^\pm T}$
denotes the produced number of $T-\ell^\pm$ correlations, where $T$ and
primary  $\ell^\pm$ originate from different $B$ mesons,
\begin{equation}
N_{\ell^+ T} \sim B(B^- \rightarrow TX) + (1-2\chi ) B(\overline B_d
\rightarrow TX)
+ 2\chi \;B(B_d \rightarrow TX) \;.
\end{equation}
\begin{equation}
N_{\ell^- T} \sim B(B^+ \rightarrow TX) + (1-2\chi) B(B_d \rightarrow
TX)
+ 2\chi \;B(\overline B_d \rightarrow TX) \;.
\end{equation}
The above two equations hold for each momentum bin of $T$,
separately.
For the dilepton analyses, $T=e^-$ and the primary component
satisfies
\begin{equation}
B(B^- \rightarrow TX) =B(\overline B_d \rightarrow TX) =B(\overline
B\rightarrow
Xe^- \bar\nu )\equiv B(b)
\end{equation}
If the secondary component would have satisfied
 \begin{equation}
B(\overline B_d \rightarrow \overline TX) = B(B^- \rightarrow
\overline TX) =
B(\overline B\rightarrow Xe^+ \nu ) \;,
\end{equation}
then we would recover Eqs.~(5.1)-(5.2), and the removal of
$B_d -\overline B_d$
mixing effects would not have to be modified.
The predominant source of $e^+$ in $\overline B$ decays originates
via the decay
chain $\overline B\rightarrow D\rightarrow e^+$.
We thus predict that
\begin{equation}
c\equiv \frac{B(\overline B_d \rightarrow e^+ X)}{B(B^- \rightarrow
e^+ X)} > 1 \;,
\end{equation}
because
\begin{equation}
B(D^+ \rightarrow e^+ X)/B(D^0 \rightarrow e^+ X) \approx \tau (D^+
)/\tau (D^0 )=
2.55\; .
\end{equation}
 The dilepton analysis requires the following modifications
(where the momentum dependence of the signal $T=e^-$ is implicit):

\begin{equation}
N_{\ell^+ T} \sim 2[(1-\chi ) B(b) +\chi c\;B(c)] \;,
\end{equation}

\begin{equation}
N_{\ell^- T} \sim (1+c-2c\chi ) B(c) + 2\chi B(b) \;.
\end{equation}
$B(b)$ denotes the primary lepton spectrum in $B$ decays,
and $B(c)$ denotes the secondary lepton spectrum from \underline{charged} $B$
decays
\cite{dssubtract},
$$B(c) \equiv B(B^- \rightarrow D\rightarrow e^+ )\;.$$
Note that the original dilepton analyses are recovered for $c=1$.
However, for the estimated value of $c\approx 1.76 $ \cite{cest} the
result is to decrease  the published CLEO II value of
\begin{equation}
B(\overline B\rightarrow Xe^-\bar\nu )=(10.49\pm 0.46)\%
\end{equation}
to \cite{wangthank}
\begin{equation}
B(\overline B\rightarrow Xe^-\bar\nu )=(10.4 \pm 0.5)\%\;,
\end{equation}
which is our estimate for this effect.

\subsubsection{Inclusive Lepton Analyses at Higher Energy Colliders}

At energies above  the $\Upsilon (4S)$ it is also possible to
determine the inclusive semileptonic BR of $B$ hadrons. The difference with
$\Upsilon (4S)$ factories
is that there are now more $B$ species being produced. At $Z^0$
factories, the production fractions [denoted by $p_i$] are
approximately
\begin{equation}
\overline B_d : B^- :\overline B_s : \Lambda_b \approx 0.4 : 0.4 : 0.12 : 0.08
\;.
\end{equation}
Thus the inclusive semileptonic BR measurements are a weighted sum over all
produced weakly decaying $B$ hadron species. Measurements at the
$Z^0$ resonance are comparable in accuracy to those at the $\Upsilon (4S)$.
A recent LEP/SLC review determined the primary component to be
\cite{lepew96}
\begin{equation}
B(b\rightarrow \ell^- )= (11.11 \pm 0.23)\% \;.
\end{equation}
This is significantly larger than the $\Upsilon (4S)$ measurements and
appears puzzling at first sight, especially since one would expect
the smaller $\Lambda_b$ lifetime to result in a smaller BR relative to
that measured at $\Upsilon (4S)$ \cite{browderh95,neubert}.
If no lifetime cuts are imposed upon the collected semileptonic
data sample, then
\begin{eqnarray}
B(b\rightarrow \ell^- ) & = & p_d \frac{\Gamma (\overline B_d
\rightarrow X
\ell^- \bar\nu )}{\Gamma (B_d)} +p_u \frac{\Gamma (B^- \rightarrow
X\ell^-\bar\nu )}{\Gamma (B^- )} + \nonumber \\
& + & \frac{p_s}{2} \;\left(\frac{\Gamma (\overline B_s \rightarrow
X\ell^-\bar\nu )}{\Gamma (B^H_s )} +\frac{\Gamma (\overline B_s
\rightarrow
X\ell^- \bar\nu )}{\Gamma (B^L_s )}\right) +p_{\Lambda_b}
\;\frac{\Gamma
(\Lambda_b \rightarrow X\ell^- \bar\nu )}{\Gamma (\Lambda_b )} \;.
\end{eqnarray}
Note that the heavy and light $B_s$ mesons could have a sizable width
difference \cite{deltagamma,bbd}.
The average $B_s$ width is however predicted to be $\Gamma (B_d)$ to
excellent accuracy \cite{bigistone,neuberts,bbd},
\begin{equation}
\frac{\Gamma (B^H_s )+ \Gamma (B^L_s )}{2} = \Gamma (B_d)
\;\left[1+{\cal O}
\left( 1\%\right)\right] \;.
\end{equation}
In addition, contrary to the widely held belief that $\tau (B^- )/\tau
(B_d)$ is larger than one \cite{bigistone}, it has been found that
theory could accommodate shorter lived $B^-$ than $\overline B_d$
\cite{neuberts,bbd}.

Great care is exercised by the LEP/SLC experiments to guarantee an
unbiased $B$ data sample for use in the extraction of $B(b\rightarrow \ell)$.
Experimentalists are aware that the extraction will be biased towards
larger $B(b\rightarrow \ell)$ values if a lifetime cut is employed
in the hemisphere of the signal lepton tag.
The longer lived $B$ species have larger inclusive
semileptonic BR under the assumption that the semileptonic decay
width is the same for all $B$ flavored hadrons.
Lifetime cuts are effective in highly enriching the $B$ data sample
and suppressing the $Z\rightarrow c\bar c \rightarrow \ell$ background.

For this reason the ALEPH collaboration prepares a pure $B$ sample by
means of requirements applied to the collection of
tracks belonging to one hemisphere in the event, and extracts
$B(b\rightarrow \ell)$
and $B(b\rightarrow c\rightarrow \ell)$ from signals in the opposite
hemisphere. This is referred to as the single lepton and same side dilepton
method \cite{alephslincl}.
The ALEPH values \cite{alephslincl} for $B(b\rightarrow \ell)$ from
this method and also the single lepton and opposite side dilepton method
need to be updated, since they both rely on an older measured momentum
spectrum  of $b\rightarrow \ell$ at threshold
machines (CLEO)~\cite{henderson}.   In particular,
a new CLEO measurement has become available~\cite{wang} which
shows a stiffer primary lepton momentum spectrum than was seen in earlier CLEO
results~\cite{henderson}.  In addition, the extraction of the primary
$b \to \ell$  spectrum from dilepton analyses at threshold machines has yet to
remove the $\overline B \to \overline D \to \ell^-$ component.
The correct extraction of the primary momentum spectrum is best
performed by CLEO, and therefore we will not attempt to quantify the
changes for $B(b \to \ell)$ at $Z^0$ factories for the above two
methods.

On the other hand, ALEPH employs the charge correlation method to
determine $B(b \to \ell)$ with no dependence on spectra obtained from lower
energy data~\cite{alephslincl}.  This allows us to be more
quantitative, especially because the extraction of $B(b\rightarrow
\ell^-)$ depends strongly on the
correct modelling of the $b\rightarrow \bar cs \rightarrow \ell^-$
background. The ALEPH collaboration recently reported a preliminary
result using this method \cite{alephslincl},
\begin{equation}
B(b\rightarrow \ell^-) = (11.01 \pm 0.38)\%\;
\end{equation}
with the $b\rightarrow \bar cs\rightarrow \ell^-$ background modelled by
\begin{equation}
B(b\rightarrow \bar cs\rightarrow \ell^- )=(1.440 \pm 0.288 )\%\;.
\end{equation}
At $\Upsilon (4S)$ factories the opposite sign dilepton data sample
has a large contribution from single $B$ decays. The CLEO collaboration
efficiently suppresses that background by the diagonal cut invented by Wang
\cite{wang}. The cut has the desirable feature of reducing the sensitivity of
$B(\overline B\rightarrow Xe^- \bar\nu )$ to the precise value of
$B(b\rightarrow  \bar cs \rightarrow e^- )$.
The central value behaves quantitatively as follows [see
Eqs.~(5.7)-(5.8) and (5.17)].
\begin{eqnarray}
B(\overline B\rightarrow Xe^- \bar\nu ) & = & 0.1095 -0.494
\;B(b\rightarrow \bar cs\rightarrow e^- )|_{\overline B\rightarrow D^-_s X} +
\nonumber \\
& - & 0.494 \cdot s \;B(b\rightarrow \bar cs \rightarrow e^-
)|_{\overline
B\rightarrow D\overline D \;\overline K X} \;,
\end{eqnarray}
where $s$ is defined just below Eq.~(5.8).
In contrast, at the $Z^0$ resonance the two $B$ hadrons generally
decay in opposite hemispheres so that no such cut is
required. In addition, the leptons experience a significant boost.
We therefore expect a larger sensitivity to
the $b\rightarrow \bar cs \rightarrow \ell^-$ background,
and indeed, evidence for this is found in the preliminary
analysis of ALEPH \cite{tenchini}
\begin{equation}
B(b\rightarrow \ell^- ) = 0.1179 - 0.54 \cdot B(b\rightarrow \bar
cs\rightarrow \ell^- ) \;.
\end{equation}
Since the $J/\psi$ and $\psi^\prime$ backgrounds already have been
explicitly subtracted by ALEPH, what remains is
\begin{eqnarray}
B(b\rightarrow \bar cs \rightarrow \ell^- ) & = & B(\overline
B\rightarrow D^-_s X) B(D^-_s \rightarrow \ell^- X) + \nonumber \\
& + & B(\overline B\rightarrow \overline D X) \;B(\overline
D\rightarrow \ell^-
X) + B(\overline B\rightarrow \overline\Lambda_c X)
\;B(\overline\Lambda_c
\rightarrow \ell^- X) \;.
\end{eqnarray}
The actual values for $B(D^-_s \rightarrow \ell^- X)$ differ for
$\Upsilon (4S)$  and $Z^0$ analyses.
The latter experiments record a larger fraction of the leptons in the
decay chain
\begin{eqnarray}
\overline B\rightarrow XD^-_s [\rightarrow && \tau^- \bar\nu ]
\nonumber \\
&& \subdecay{\ell^- \nu\bar\nu}
\end{eqnarray}
than is true for the former experiments, because of the large boost of the
$B$ hadrons.
Whereas CLEO/ARGUS used only $e^\pm$ as signal leptons, the LEP/SLC
uses both $e^\pm$ and $\mu^\pm$.
[ALEPH assumes the same background BR for electrons and
muons~\cite{alephslincl}. We caution that the  process $B(D^-_s
\rightarrow \mu^- \bar\nu )\approx 1\%$ enhances the
$b\rightarrow \bar cs \rightarrow \mu^-$ background over that of the
$b\rightarrow \bar cs \rightarrow e^-$.]
To make our point more forcefully, we oversimplify and almost ignore
process (5.39) for CLEO/ARGUS while taking it fully into account for $Z^0$
factories. The $D_s$'s from $\overline B\rightarrow D^-_s X$ then satisfy :
\begin{eqnarray}
B(D^-_s \rightarrow \ell^- X)|_{Z^0} & = & B(D^-_s \rightarrow e^-
X)|_{\Upsilon (4S)} + \nonumber \\
& + & B(D^-_s \rightarrow \ell^- \bar\nu ) +B(D^-_s \rightarrow
\tau^-\bar\nu )
\;B(\tau^- \rightarrow \ell^- \nu\bar\nu ) \approx \nonumber \\
& \approx & 6.64 \times 10^{-2} + 0.0091 + 0.091 \times 0.18 = 0.092
\; .
\end{eqnarray}
Here the values are for the case of a muon. The background is then estimated as
\begin{equation}
B(b\rightarrow \bar c s \rightarrow \mu^- ) = 0.10 \times 0.092 +
0.01 +
1.6\times 10^{-4} \approx 0.02 \;.
\end{equation}
This larger background reduces the primary lepton BR of ALEPH to
\begin{equation}
B(b\rightarrow \ell^- )= (10.7 \pm 0.4 )\%\;.
\end{equation}
Clearly, the correct acceptances and efficiencies for each relevant
process must be obtained by the various $Z^0$ experiments.
One of the main points of this subsection is that now that CLEO
has completed the mapping out of the $b\rightarrow c+\bar cs$ processes, we
recommend  the use of the measured BR's and momentum spectra of the
wrong-charm
$b\rightarrow \bar c$ transitions for a correct modelling of the
$b\rightarrow \bar  cs
\rightarrow \ell^-$ background at the $Z^0$.  The removal of $B^0 -
\overline B^0$ mixing effects is more subtle than currently performed
by ALEPH.

The secondary lepton component experiences different (probably
larger) $B - \overline B$ mixing effects than the primary lepton
component (see Section V.A.2.b).  The recent charge correlation
method presented by ALEPH did not take into account different mixing effects
of primary and secondary leptons.  This is not the only analysis that has
to be modified for such effects.  All published
reports~\cite{mixingwrong}, which determine the average
(time integrated) mixing parameter $\overline \chi$ from dilepton
analyses, must be modified, because they implicitly assumed the same
average mixing effects for the primary and secondary lepton
components. This impacts measurements of $B(b \to \ell^-)$ that
involve $\overline \chi$.

We hope to have motivated experimentalists to reanalyze their data so
as to find out the cause of the apparent puzzle of a significantly
larger  inclusive semileptonic BR of $B$ hadrons at the $Z^0$
resonance than at the $\Upsilon(4S)$.

\subsection{Exclusive Semileptonic $B$ Decays}

Table VI catalogues the various exclusive semileptonic processes.
Class 1 consists of the exclusive $\overline B\rightarrow D^* \ell^-\bar\nu$
processes. The most accurate measurement of such a process, where
$D^{*+}$'s are fully reconstructed, was performed by CLEO~\cite{barish,burchatr}
\begin{equation}
B(\overline B^0 \rightarrow D^{*+} \ell^- \nu )=(4.49 \pm 0.50)\%\;,
\end{equation}
whereas that for the charged $B$ decay has a larger error~\cite{burchatr},
\begin{equation}
B(B^- \rightarrow D^{*o} \ell^- \bar\nu ) = (5.34 \pm 0.80)\% \;.
\end{equation}
The last two equations used the 1994 PDG  values for
BR's of the weakly decaying charm decays \cite{burchatr}. We
will be conservative and keep the full error and factor out
$B(D^0  \rightarrow K^- \pi^+ )$ explicitly (see Table VI).
Most recently CLEO~\cite{kutschke} reported accurate measurements for
exclusive $\overline B^0 \to D^+$
transitions via the missing mass and neutrino reconstruction
techniques, respectively,
\begin{eqnarray}
B(\overline B^0 \rightarrow D^+ \ell^- \overline \nu ) \;\; B(D^+
\rightarrow K^-\pi^+ \pi^+)& = &
0.00159 \pm 0.00029 \; \;, \nonumber \\
 & = & 0.00172 \pm 0.00036\;\;\;\;\;.
\end{eqnarray}
The exclusive $ B^- \to D^0$ BR measurement of CLEO has a larger
error, namely,
\begin{equation}
B(B^- \rightarrow D^0 \ell^- \overline \nu )  = (1.95 \pm
0.55)\%\;\;\;\;\cite{drellp}\;.
\end{equation}

Since the experimental situation regarding $p$ wave states is
controversial
(ARGUS \cite{arguspwave} and OPAL \cite{opalpwave} claim to see most
of
the remainder as
$p$ wave excitations, whereas ALEPH \cite{alephpwave} and CLEO
\cite{drellp,bellerive} do not)
and since current experiments cannot observe semileptonic $B$ decays
with broad
charmed $p$ waves, we resort to a model independent sum rule
\cite{sumrule}.
This sum rule obtains the sum over all (both narrow and broad)
charmed $p$ wave
semileptonic BR's, $\overline B\rightarrow D^{**} \ell\nu$ from the
slope $\rho^2$  of
the Isgur-Wise function which parameterizes the $\overline
B\rightarrow D^{(*)}\ell\nu$ process,
\begin{equation}\frac{B(\overline B\rightarrow D^{**} \ell\nu
)}{B(\overline
B\rightarrow X_c \ell\nu )} \approx \frac{\left(\rho^2
-\frac{1}{4}\right)}{0.5} \;\;(0.08 \pm 0.04) \;.
\end{equation}
Current slope values yield that about 10\% of the inclusive
semileptonic decays
are $D^{**}\ell\bar \nu$ \cite{isgurprivate}, a value consistent with
calculations of various models \cite{isgw2,veselio}.

One may argue that present data rule out such a small $p$ wave
contribution \cite{arguspwave,opalpwave},
but none of the existing analyses have removed the $\overline B\rightarrow
D\overline D \;\overline K X$ background.
ARGUS \cite{arguspwave} claimed to have measured
$$ B(\overline B\rightarrow D^{**} \ell\nu ) = (2.7 \pm 0.5  \pm
0.5 ) \%\;,$$
by applying the missing mass technique to their $\overline
B\rightarrow D^*\ell^ - X$ data sample \cite{arguspwave}. But the shape of
their missing mass spectrum indicates [at least to us] a sizeable $\overline
B\rightarrow D\overline D^* \overline KX, D^* \overline D\;\overline KX$
component where one of the charms decayed semileptonically.
To buttress their claim, ARGUS searched for additional $\pi^-$'s,
which they then correlated with the $D^{*+}$, to form a peak. While the
additional $\pi^-$ has some discriminating power, there could still
be sizable backgrounds such as
\begin{eqnarray}
\overline B\rightarrow D^{*+} && D^{*-} \overline K X \nonumber \\
&& \subdecay{\pi^-  \overline D^0 [\rightarrow \ell^- X]}
\end{eqnarray}
Note that we have assumed that no $\pi^- /K^-$ misidentification has
been made. To the extent that this is not true, more background
sources become important.

The LEP measurements can use significantly displaced vertices
to isolate their $\overline B\rightarrow D^{**} \ell^- \bar\nu $
signal. If only loose particle identification cuts have been applied, then the
primary (or secondary) kaon in
$\overline B\rightarrow D^{(*)} \overline D^{(*)} \overline K X$
could have been
combined with the $D^{(*)}$ to form a fake signal. Another potential
problem occurs if a sizable fraction of the background is seen as
$$\overline
B\rightarrow D^{**} \overline D[\rightarrow \ell^- KX]   \overline KX
= D^{**} \ell^- X \;.$$
This background is removable, for instance, on account of the two
additional kaons
it generates, via differences in $D^{**}\ell^-$ correlations, or
by means of momentum or relative transverse momentum spectrum of the lepton.
Because experiments have to sort out what exactly has been measured,
we suggest for now the use of the Bjorken-Isgur-Wise sum rule.

Model calculations find negligible semileptonic BR's with excited charm
resonances beyond the $p$ wave states \cite{veselio,isgw2}. Because
also the
truly ``non resonant" $[D^{(*)} n\pi\ell\bar\nu , D_s \overline
KX\ell\bar\nu ]$ component is
predicted to be small \cite{yang,goityr,veselio,adgr} [at the
(5-10)\%
level of the inclusive
semileptonic BR], it appears that the sum over all exclusive
semileptonic modes
does not saturate the inclusive semileptonic BR (see Table VI).
To see the excess clearly, choose the recent CLEO value
\begin{equation}
B(D^0 \rightarrow K^- \pi^+)=(3.91 \pm 0.19)\% \;,
\end{equation}
and use classes $1^+, 2-5$ of Table VI. The combined BR is
\begin{equation}
BR(1^+ )+BR(2-5) =(8.2 \pm 0.9)\%
\end{equation}
and falls significantly below the inclusive measurements.
[If the more poorly measured $B(\overline B \rightarrow D^{*o} \ell^-
\bar\nu )$ is
used, one obtains from Table VI,
\begin{equation}
BR(1^0) +BR(2-5)=(9.1 \pm 1.1)\%\;.]
\end{equation}
The $\Upsilon(4S)$ experiments have an additional experimental handle
which could shed light upon this issue~\cite{4spuzzle}.

We note that the shortfall is solved by
sufficiently reducing the value of $B(D^0 \rightarrow K^- \pi^+)$. Table VI
furnishes the BR's for the equation~\cite{tnocharm}
\begin{equation}
BR(1^+) +BR(2-5) = BR(t) \;[BR(1^0) +BR(2-5) = BR(t)].
\end{equation}
The above equation is solved for the following value
\begin{equation}
B(D^0\rightarrow K^- \pi^+ )= (2.9 \pm 0.4)\% \;[(3.3 \pm 0.5)\%].
\end{equation}

There is no reason yet to be concerned  about the smallish $B(D^0
\rightarrow K^-\pi^+)$ value, because the coefficient of $(\rho^2 -
1/4)$ for the p wave BR (Class 3 of Table VI) could have been
underestimated, and so could the nonresonant BR  (Class 5 of Table
VI).  The accurate measurements of $B(\overline B\rightarrow D^{*}
\ell \overline \nu )$ and
$B(\overline B\rightarrow D \ell \overline \nu )$ allow CLEO to
perform consistency checks with the inclusive data samples.  This
enables CLEO to probe the fraction of the time semileptonic decays
are not seen as $\overline B\rightarrow D^{(*)} \ell \overline \nu$.
Future theoretical investigations would then be useful in estimating
how to partition that remainder into Classes 3 and 5 of Table VI.

There appears to exist weighty evidence from the low values of $R_c,
n_c$ and sum over exclusive semileptonic BR's of $B$ decays that $B(D^0
\rightarrow K^-\pi^+)$ ought to be lower than presently estimated.
The precise determination of $B(D^0 \rightarrow K^-\pi^+)$ is of
paramount importance, because this mode calibrates much that is known
in heavy flavor (both charm and beauty) decays.  As indicated above,
it is of significance in ascertaining whether there is a Standard
Model reason for the apparent discrepancy between theory and
experiment regarding $R_b$. The next section thus suggests several
methods that
accurately measure $B(D^0 \rightarrow K^- \pi^+ )$ from correlations
with the primary
lepton data sample from $B$ decays.  One of them uses the fact that
\begin{equation}
B(\overline B\rightarrow D^+ \ell^- X)  +   B(\overline B\rightarrow
D^0
\ell^- X) + B(\overline B\rightarrow D^+_s \ell^- X) \approx
B(\overline
B\rightarrow X_c
\ell^-)\;.
\end{equation}
This method does not involve soft $\pi^+$ detection
efficiencies [from
$D^{*+} \rightarrow \pi^+ D^0$ decays], and is thus complementary to
the existing determinations of $B(D^0 \rightarrow K^-\pi^+)$ \cite{pdg}.

\section{Measuring $B(D^0 \rightarrow K^- \pi^+ )$ from $\overline
B\rightarrow
X_c \ell\nu$ Processes}

Our suggestion that a sizable fraction of all $B$ decays are seen as
$\overline B\rightarrow D\overline D \;\overline KX$ \cite{bdy}
has recently been proven by CLEO from $\ell^\pm -D$ correlations
\cite{moriond}.
A severe momentum cut $(p> 1.5$ GeV/c) guarantees the lepton to be primary, and
angular correlations between $\ell - D$ allow us to measure
separately the two cases in which
\cite{tipton}:
\begin{enumerate}
\begin{enumerate}
\item the lepton originates from one $B$ and the $D$ originates from the
other $B$ in the event, and
\item both the lepton and the $D$ come from a single $B$ decay.
\end{enumerate}
\end{enumerate}
Whereas case (a) probes our suggestion, case (b) is important in its
own right because it provides an accurate method for determining
$B(D^0 \rightarrow K^- \pi^+ )$:

The number of $\overline B\to$ $D^0 \ell^- ,D^+ \ell^- ,D^+_s \ell^- ,
\Lambda^+_c \ell^-$ must equal the number of $\ell^-$
produced from $b\rightarrow c\ell^-$ processes.    Here all leptons
satisfy a high momentum cut of say $p> 1.5$ GeV/c.
Backgrounds from the continuum and from $b\rightarrow u\ell^-$
transitions have been subtracted.
Good vertexing would allow further suppression of the backgrounds
relative to the
$b\rightarrow c\ell^-$ signal.
 The semileptonic decay $\overline   B\rightarrow
D^+_s \overline K X\ell\nu$ has not yet been observed, and stringent
upper
limits   exist \cite{chomenary}.
While it is expected to occur with a small few permille BR to start
with, it will
be even less important fractionally for the high end lepton momentum
region
advocated here. This process is currently being estimated
\cite{adgr}.
The process $\overline B\rightarrow \Lambda^+_c \overline NX\ell^-
\bar\nu$ is
expected to be utterly negligible at the current level of accuracy.
 Denote the
produced number of events $(f\ell^-)$ from $\overline B \rightarrow
X_c \ell^- \bar\nu$ transitions by
$N[f\ell^-]$. Then
\begin{equation}
\frac{N[D^0 \ell^- ] +N[D^+\ell^- ] + N[D^+_s \ell^- ]+
N[\Lambda^+_c \ell^-]}{N[c \;\ell^-]} =1 \;,
\end{equation}
where $N[c \;\ell^-]$ denotes the produced number of $b\rightarrow
c\ell^-$ processes.
Define further,
\begin{equation}
L\left[\left(f\right)_H\right]\equiv\frac{N\left[\left(f\right)_H\ell^
-
\right]}{N[c \;\ell^-]}\;,
\end{equation}
\begin{equation}
r_+ \equiv \frac{B(D^+\rightarrow K^- \pi^+ \pi^+ )}{B(D^0
\rightarrow K^-
\pi^+)} \;, \;\;r_s \equiv \frac{B(D^+_s\rightarrow \phi \pi^+
)}{B(D^0
\rightarrow K^- \pi^+)}\;\;\; .
\end{equation}
The desired BR is then
\begin{equation}
B(D^0 \rightarrow K^- \pi^+) = \frac{
L[(K^- \pi^+)_{D^0}] + \;L[(K^- \pi^+ \pi^+ )_{D^+}]\;/\;r_+\;+
\;L[(\phi\pi^+
)_{D_s}] /\;r_s\;}
{1-L[(pK^- \pi^+ )_{\Lambda_c}] \;
\frac{1}{B(\Lambda_c \rightarrow pK^- \pi^+)}
} \;.
\end{equation}
$B(D^0 \rightarrow K^- \pi^+)$ is measured by
determining the quantities on the right hand side of the above
equation, where,
for instance, $L[(K^- \pi^+)_{D^0}]$ denotes the fraction of
$\overline
B\rightarrow (K^- \pi^+)_{D^0}\;X\; \ell^-\bar\nu$ events relative to
$\overline B\rightarrow X_c \ell^- \bar\nu$ events~\cite{kwon}.

For now the use of 1 for the denominator is a useful approximation,
and the $D^+_s \ell^- /(c\: \ell^-)$
fraction could either be taken from theory \cite{adgr} or from
experiment once it is observed.
Since the $D^+ \ell^- /(c\: \ell^-)$ fraction will be substantially
smaller than $D^0
\ell^- /(c\: \ell^-)$ [because of isospin violating $D^*$ decays],
the
ratio of BR's $r_+$
need not be known to the same degree of accuracy that is pursued for
$B(D^0\rightarrow K^- \pi^+ )$.
Still, Menary suggested that it may be feasible to determine $r_+$ to
2\% accuracy from the $\Upsilon (4S)$ continuum \cite{menaryrp}.
To augment statistical data, we suggest the use of all available
final states of
charmed hadron decays, because the ratios of BR's to the calibrating
modes are well known.

With straightforward modifications, higher energy experiments (such as $Z^0$ factories) may
wish to study the feasibility of determining $B(D^0 \rightarrow K^- \pi^+ )$ from
an analogous analysis.  This determination of $B(D^0 \rightarrow K^- \pi^+)$ is complementary to
the present measurements of $B(D^0 \rightarrow K^- \pi^+ )$, in that
it does not
require the observation of the soft $\pi^+$ from $D^{*+} \rightarrow
\pi^+ D^0$ decays.

The process $\overline B_d \rightarrow D^{*+} \ell^-\bar\nu$ allows a
second method for determining $B(D^0 \rightarrow K^- \pi^+)$~\cite{argusdstar}. Here one
infers the existence of a $D^{*+}$ from its soft $\pi^+$ daughter. CLEO
\cite{yamamotod0kpi} could extract $B(D^0 \rightarrow K^- \pi^+)$ by
measuring the fraction of the time where the $\pi^+ \ell^-$ sample involves an
additional $(K^- \pi^+)_{D^0}$, which gives the $D^{*+}$ peak when combined
with the soft $\pi^+$.

Hadron facilities (such as the CDF collaboration), $Z^0$ factories
and asymmetric $B$ factories could suppress backgrounds by requiring
the $\pi^+ \ell^-$
sample to form a detached vertex. Once this sample is correlated with a fully
reconstructed $D^0$, which together with the soft $\pi^+$ forms the
$D^{*+}$, $B(D^0 \rightarrow K^- \pi^+)$ can be extracted.
In an analogous spirit, the mode $\overline B_d \rightarrow D^{*+}
\pi^-$ could be also used \cite{yamamotod0kpi}.

We hope to have convinced the reader of the importance of the $D^0
\rightarrow K^- \pi^+$ BR, and the need for additional high precision
measurements.

\section{Heavy Baryon Production and Decay}

The calibrating mode of heavy baryon decays is $\Lambda_c\rightarrow pK^- \pi^+$.
While the world average is \cite{pdg}
\begin{equation}
B(\Lambda_c \rightarrow pK^- \pi^+ ) = (4.4 \pm 0.6)\% \;,
\end{equation}
this section argues to use instead \cite{shipsey}
\begin{equation}
B(\Lambda_c \rightarrow pK^- \pi^+ )= (6.0 \pm 1.5)\%\;.
\end{equation}
The world average is dominated by baryon production analyses in $B$ decays
\cite{crawford,argusbaryon} which assumed that almost always a weakly decaying
charmed baryon is involved.
The assumption is invalidated by the observation of significant $\overline B$
decays into a charmed meson and a nucleon and/or anti-nucleon,
\begin{equation}
\overline B\rightarrow D^{(*)} NX, D^{(*)} \overline N X \;.
\end{equation}
A straightforward Dalitz-plot analysis predicts that a sizable fraction of all
baryon producing $\overline B$ decays is likely to be of that form (see the
Appendix).
The $b\rightarrow c+W$ transition, where the virtual $W$ hadronizes as a
baryon-antibaryon pair plus perhaps additional debris, provides another source
for $\overline B\rightarrow D^{(*)}NX, D^{(*)} \overline NX$ processes. Flavor
correlations within the $\overline B\rightarrow D^{(*)} \stackrel{(-)}{N} X$
processes allow to distinguish between the two production mechanisms.

An extension of the $\ell^+D^{(*)}$ correlations presented recently by CLEO
\cite{moriond}, may yield the first observation of such processes. A positively
charged high momentum lepton and a $D^{(*)}$ meson must originate from two
different $B$'s in the $\Upsilon (4S)$ event. An additional nucleon or
antinucleon should be searched in that data sample. A positive signal would then
most likely demonstrate the existence of the processes given in Eq.~(7.3), or very
unlikely baryon production in semileptonic $\overline B$-decays
\cite{yamamotobaryon}. The existence of angular correlations between the
$\stackrel{(-)}{N}$ and $D^{(*)}$ would then prove that both particles originate
from the same $B$ \cite{thorndike}.
If such correlations are found to be sizable (which we predict), the 1994 and
1996 PDG world average for $B(\Lambda_c\rightarrow pK^-\pi^+)$ must be abandoned
in favor of \cite{shipsey}
$$B(\Lambda_c\rightarrow pK^-\pi^+) = (6.0\pm 1.5)\%\;.$$
One consequence would be that current measurements overestimate heavy baryon
production.

\section{Implications and Conclusions}

When we try to make sense of the combined experimental evidence
reviewed here, we conclude that the absolute BR of charmed hadrons must be
reevaluated. This note
considered the evidence from the  (a)  $n_c$ values in $B$
decays, (b) $R_c$ measurements at
the $Z^0$ resonance, and (c) the combined exclusive semileptonic
$B$ decay BR's versus the inclusive semileptonic BR.
We identified $D^0 \rightarrow K^- \pi^+$ as
the main culprit and expect that
its BR will decrease from the presently accepted value.
Some of the consequences of this are that
\begin{enumerate}
\begin{enumerate}
\item a sizable fraction of $D^0$ decays may still have to
be accounted for;
\item $B(\overline B\rightarrow D^{(*)} \ell\bar\nu )$ will
increase, causing its derived $|V_{cb}|$ to increase;
\item $B(\overline B \rightarrow
\stackrel{\textstyle{(-)}}{\textstyle{D^0}} X)$ and $B(\overline
B\rightarrow
D^\pm X)$ will increase;
\item $B(D_s \rightarrow \phi\pi ^+)$ would decrease, because of
its recent ``model independent" extraction which relates $B(D_s
\rightarrow \phi\pi^+ ) \sim B(D^0 \rightarrow K^- \pi^+ )$ \cite{cleods};
and  thus
\item $B(\overline B\rightarrow D^\pm_s X)$ would increase.
\end{enumerate}
\end{enumerate}
The $\overline B\rightarrow D\overline D \;\overline K X$ processes
were neglected in all existing experimental analyses. But CLEO has
demonstrated that
they have a sizable BR \cite{bdy}, with some of the following
consequences:
\begin{enumerate}
\begin{enumerate}
\item the current understanding of primary and secondary lepton
spectra in
$B$ decays has to be modified;
\item $B(\overline B\rightarrow X_c \ell \bar\nu )$ will have to
be modified;
\item the exclusive $B^-$ and $\overline B_d$ lifetime determinations
from $\overline B\rightarrow D^{(*)} X\ell^- \bar\nu$ data samples will
have to be modified; and
\item the traditional belief that $\overline B\rightarrow D^{(*)}
X\ell^-
\bar\nu$ processes unambiguously tag their parent $B$ flavor at time
of decay is
not true, because of the following background
\begin{eqnarray}
\overline B\rightarrow && D\overline D \;\overline K
X=\ell^+ \overline D  \;\;\;. \nonumber \\
&& \subdecay{\ell^+} \nonumber
\end{eqnarray}
\end{enumerate}
\end{enumerate}
The discovery and accurate modeling of $B$ flavor tags is crucial for $B^0_q
-\overline B^0_q$ mixing ($q=d$ or $s)$ and CP violation studies.
Thus we note
that the $\overline B\rightarrow D\overline D \;\overline K X$
background is
removable from the semileptonic $\overline B\rightarrow DX\ell^-
\bar\nu$
processes on several accounts:
\begin{enumerate}
\item softer lepton spectrum (in $p$ and $p_{T,re\ell}$);
\item different $D^{(*)}-\ell$ correlations; and
\item the two additional kaons generated in background events:
Primary kaon from
the $b\rightarrow c\bar cs$ transition, and secondary kaon from
semileptonic
charm decay ($c\rightarrow s\ell^+ \nu \;{\rm or}\; \bar c\rightarrow
\bar
s\ell^- \bar\nu)$.
\end{enumerate}
While this background can be accurately taken into account
and is easily
removable, none of the present experiments have done so. We are eager
to learn what will
change in the published measurements, once the effects in this note
have been included.

\section{Acknowledgements}

We thank M.S. Alam, J. Appel, T. Behnke, M. Beneke, K. Berkelman, D. Bloch, G. Buchalla, D. Charlton,
P.S. Cooper, J. Cumalat, A.S. Dighe, P. Drell,
E. Eichten, R. Forty, A. Freyberger, D. Fujino, E. Graziani, J. Incandela, N.
Isgur, V. Jain, R. Kowalewski, R. Kutschke, Y. Kwon, J.D. Lewis, A.J. Martin,
R. Poling, J. Richman, T. Riehle, A. Ryd, D. Scora, F. Simonetto, R. Snider, R.
Tenchini, E. Thorndike,
S. Veseli, R. Wang,  and H. Yamamoto for discussions.
We are most grateful to G. Hockney for writing an error propagation
package for us; to J. Incandela for double checking the
error propagation, for many lively discussions and for help in
focussing the present paper;
and to R. Wang for many useful discussions and for modifying his
lepton data sample to study the various effects discussed in this
paper.  We are grateful to Vincent Whelan for useful comments on an
earlier draft, and to Gerhard Buchalla and Joe Incandela for comments on a later
version.  Many thanks go to Lois Deringer who diligently typed up this manuscript
and many earlier drafts.
This work was supported in part
by the   Department
of Energy, Contract No. DE-AC02-76CH03000.

\section{Appendix: (Charmed) Baryon Production in $B$ decays}

The accurate accounting of inclusive charm yields in $B$ decays
requires a
consistent description of charmed baryon production which is
lacking in the
existing literature. Two years ago we hypothesized that the soft
inclusive
momentum spectrum of inclusive $\Lambda_c$ production indicates that
$b\rightarrow c\bar cs$ is the dominant source of $\Lambda_c$'s in
$B$ decays \cite{dcfw}. Our hypothesis predicted (i) large wrong-sign $\ell^-
\Lambda_c$ correlations, where the lepton comes from the semileptonic decay of
one $B$ and the $\Lambda_c$ from the other $B$ in an $\Upsilon (4S)$ event;
and (ii) large $\Xi_c$ production in $B$ decays, which at the time had not
been observed and was believed to be greatly suppressed~\cite{crawford}.
Within two months, CLEO observed the first evidence of $\Xi_c$
production in $B$ decays (see Table I), but proved that the right-sign
$\ell^+\Lambda_c$
correlations are dominant (see Table III) \cite{glasgowbaryon}.
Removing $B^0 -\overline B^0$ mixing effects as outlined in
Ref.~\cite{distinguish}, CLEO measured
\cite{glasgowbaryon}
\begin{equation}
r_{\Lambda_c} \equiv \frac{B(\overline B\rightarrow
\overline\Lambda_c
X)}{B(\overline B\rightarrow \Lambda_c X)} = 0.20 \pm 0.14 \;,
\end{equation}
where the error does not include a possible systematic uncertainty
emphasized in \cite{distinguish} and mentioned in Section V.A.2.b above.

Because the CLEO measurements of inclusive $\Xi_c$ production in $B$ decays
involve large uncertainties and their central values appeared to us
to be too high, this appendix correlates the $\Xi_c$ and $\Omega_c$ production
in tagged
$\stackrel{(-)}{B}$ decays to that of the $\Lambda_c$ which has been
measured with greater
accuracy. We neglect $b\rightarrow u$ transitions and use the Cabibbo
suppression factor $\theta^2 = (0.22)^2$ for charmed baryon
production in
$b\rightarrow c\bar us (b\rightarrow c\bar cd)$ versus $b\rightarrow
c\bar
ud^\prime (b\rightarrow c\bar cs^\prime )$ transitions.
The parameter $p=0.15 \pm 0.05$ models $s\bar s$ fragmentation
relative to $f\bar f$ fragmentation from the vacuum, where $f=u,d$ or
$s$. The large numerical value for $p$ was chosen on purpose.  It is used
to demonstrate that even large values of $p$ yield a
significant reduction in $\Xi_c$ production in $\overline B$ decays.

Denote by $C_{\bar ud}$ the fraction of $\overline B$ decays to weakly
decaying charmed baryons which come from $b\rightarrow c\bar ud$,
and define $C_{\bar cs}, C_{\bar cd}, C_{\bar us}$ analogously. Because our
model allows
for substantial charmless-baryon charmless-anti-baryon production in
$B$ decays, $C_{\bar
ud}$ is smaller or at most equal to $B_{\bar ud}$ defined in
Ref.~\cite{dcfw}.
Similar comments hold for the analogous $C$ and $B$ quantities.
The simplest version of the model predicts
\begin{equation}
B(\overline B\rightarrow \Lambda_c X) = (1-p)(C_{\bar ud} + C_{\bar
cd})
\end{equation}
\begin{equation}
B(\overline B\rightarrow \overline\Lambda_c X) = (1-p) (C_{\bar cs} +
C_{\bar
cd})
\end{equation}
\begin{equation}
B(\overline B\rightarrow \Xi_c X) = p\; C_{\bar ud} + (1-p)\; C_{\bar
us} + (1-p)\;
C_{\bar cs} + p\; C_{\bar cd}
\end{equation}
\begin{equation}
B(\overline B\rightarrow \overline\Xi_c X) = p\; (C_{\bar cs} +
C_{\bar cd})
\end{equation}
\begin{equation}
B(\overline B\rightarrow \Omega_c X) = p(C_{\bar us} + C_{\bar cs})
\end{equation}
\begin{equation}
B(\overline B\rightarrow \overline\Omega_c X) = 0
\end{equation}
The Cabibbo structure
\begin{equation}
C_{\bar cd}/(C_{\bar cd} +C_{\bar cs})= C_{\bar cd}/C_{\bar
cs^\prime} = \theta^2
\end{equation}
\begin{equation}
C_{\bar us}/(C_{\bar us}+C_{\bar ud}) = C_{\bar us}/C_{\bar
ud^\prime} =
\theta^2
\end{equation}
allows one to express the six observables listed on the left hand
sides of
Eqs.~(10.2)-(10.7) in terms of the two unknowns $C_{\bar ud}$ and
$C_{\bar cs}$. These are obtained, in turn, from the two
measurements involving inclusive $\Lambda_c$ production in $B$ decays,
namel $Y_{\Lambda_c}$ and $r_{\Lambda_c}$, as follows~:

\begin{equation}
\frac{C_{\bar ud}}{Y_{\Lambda_c}}=\frac{(1+\lambda^2 -\lambda^2
r_{\Lambda_c})}{(1-p) (1+\lambda^2 )(1+r_{\Lambda_c})}\;,
\end{equation}
\begin{equation}
\frac{C_{\bar cs}}{C_{\bar ud}} = \frac{r_{\Lambda_c}}{1+\lambda^2
(1-r_{\Lambda_c})} \; ,
\end{equation}
where
\begin{equation}
\lambda^2 =\frac{\theta^2}{|V_{cs}|^2} =
\frac{\theta^2}{\left(1-\frac{1}{2}
\theta^2 \right)^2} \; .
\end{equation}
The inclusive $\stackrel{(-)}{\Xi_c},
\stackrel{(-)}{\Omega_c}$ yields in $\overline B$ decays are thus
correlated to inclusive $\stackrel{(-)}{\Lambda_c}$ production,
\begin{equation}
\frac{B(\overline B\rightarrow \Lambda_c X)}{Y_{\Lambda_c}} =
\frac{1}{1+r_{\Lambda_c}} \;,
\end{equation}
\begin{equation}
\frac{B(\overline B\rightarrow \overline\Lambda_c
X)}{Y_{\Lambda_c}}=\frac{r_{\Lambda_c}}{1+r_{\Lambda_c}} \;,
\end{equation}
\begin{equation}
\frac{B(\overline B\rightarrow \Xi_c X)}{Y_{\Lambda_c}}=\frac{C_{\bar
ud}}{Y_{\Lambda_c}} \left\{p+(1-p)\lambda^2 +\frac{C_{\bar
cs}}{C_{\bar ud}}
(1-p+p\lambda^2 )\right\}\;,
\end{equation}
\begin{equation}
\frac{B(\overline B\rightarrow \overline\Xi_c
X)}{Y_{\Lambda_c}}=\frac{C_{\bar
ud}}{Y_{\Lambda_c}} \;\frac{C_{\bar cs}}{C_{\bar ud}} \;p(1+\lambda^2
)\;,
\end{equation}
\begin{equation}
\frac{B(\overline B\rightarrow \Omega_c X)}{Y_{\Lambda_c}} = p
\frac{C_{\bar
ud}}{Y_{\Lambda_c}} \left(\lambda^2 +\frac{C_{\bar cs}}{C_{\bar
ud}}\right) \;,
\end{equation}
\begin{equation}
B(\overline B\rightarrow \overline\Omega_c X) = 0 \;.
\end{equation}

We have taken $p$ to be a universal quantity and have assumed that
the initially produced charmed baryon retains its charm [and when
applicable, its strange] quantum number[s] through to its
weakly decaying offspring. That is not justified.
We typically expect the initially produced charmed baryons (via
$b\rightarrow
c)$ to be highly excited, while this is not expected of their
pair produced
antibaryons (via $b\rightarrow \bar u$ or $b\rightarrow \bar
c)$~\cite{dalitz}.
That scenario explains naturally the puzzling soft momentum spectrum of the inclusive
$\Lambda_c$ yield in $\overline B$ decays~\cite{zoeller}.
That a sizable fraction of these highly excited charmed baryons could
break up
into a charmed meson, a charmless baryon and additional debris is
irrelevant to
our discussion which focusses on weakly decaying charmed baryon
production in $B$ decays~\cite{tightlimit}.
In contrast, it is important to note that $\Xi^r_c \rightarrow
\Lambda_c \overline K X$ could occur significantly [where the superscript
$r$ denotes excited resonances], and can be tested by observing
$\Lambda_c \overline \Lambda$ correlations from single $\overline B$ decays.
This introduces an additional mechanism for $\Lambda_c$ production in
$\overline B$ decays, which may help explain the small measured value of
$r_{\Lambda_c}$. It also decreases the naive estimate for weakly
decaying $\Xi_c$ production. Because our predictions have not
incorporated such
effects, they should be viewed strictly as upper limits for $\Xi_c $
production in $\overline B$ decays.

\begin{table}
\caption{Inclusive Charmed Hadron Production in $B$
Decays as Measured by CLEO}
\begin{tabular}{|c|c|c|}
$T$ & $Y_T \equiv B(\overline B\rightarrow TX) +
B(\overline B\rightarrow \overline T X)$ & Reference \\
\tableline
$D^0$ & $(0.645 \pm 0.025) \left[\frac{3.91\%}{B(D^0 \rightarrow K^-
\pi^+)}\right]$ & \cite{yamamoto} \\
\tableline
$D^+$ & $(0.235 \pm 0.017) \left[\frac{9.3\%}{B(D^+ \rightarrow K^-
\pi^+\pi^+
)}\right]$ & \cite{yamamoto} \\
\tableline
$D$ & $(0.883 \pm 0.038) \left[\frac{3.91\%}{B(D^0 \rightarrow
K^-\pi^+)}\right]$ & \\
\tableline
$D_s$ & $(0.1211 \pm 0.0096) \left[\frac{3.5\%}{B(D_s \rightarrow
\phi\pi )}\right]$ & \cite{menary} \\
\tableline
$\Lambda_c$ & $(0.030 \pm 0.005) \left[\frac{6\%}{B(\Lambda_c
\rightarrow pK^-
\pi^+)}\right]$ & \cite{zoeller} \\
\tableline
$\Xi^+_c$ & $0.020 \pm 0.007$ & \cite{jain}\\
\tableline
$\Xi^0_c$ & $0.028 \pm 0.012$ & \cite{jain}

\end{tabular}
\end{table}

\begin{table}
\caption{Absolute Branching Ratios of Key Charm Decays as used by
CLEO}
\begin{tabular}{|c|c|c|}
Mode & BR [in \%] & Reference \\
\tableline
\tableline
$D^0 \rightarrow K^- \pi^+$ & $3.91 \pm 0.19$ & \cite{cleod0} \\
\tableline
$D_s\rightarrow \phi\pi$ & $3.5 \pm 0.4$ & \cite{pdg} \\
\tableline
$\Lambda_c \rightarrow pK^- \pi^+$ & $6.0 \pm 1.5$ & \cite{shipsey}
\end{tabular}
\end{table}

\begin{table}
\caption{Inclusive Charmed Hadron Production in Tagged $B$ Decays as
Measured by
CLEO}
\begin{tabular}{|c|c|c|}
Observable & Value & Reference \\
\tableline
\tableline
$r_{\Lambda_c} \equiv \frac{B(\overline B\rightarrow
\overline\Lambda_c X)}{B(\overline
B\rightarrow \Lambda_c X)}$ & $ 0.20 \pm 0.14$ & \cite{glasgowbaryon}
\\
\tableline
$r_D \equiv \frac{B(\overline B\rightarrow \overline DX)}{B(\overline
B\rightarrow
DX)}$ & $0.107 \pm 0.034 $ & \cite{moriond} \\
\tableline
$f_{D_s} \equiv \frac{B(\overline B\rightarrow D^+_s X)}{Y_{D_s}}$ &
$0.172 \pm 0.083$ & \cite{chomenary}

\end{tabular}
\end{table}

\begin{table}
\caption{Inclusive Semileptonic BR for Various Models from
Ref.~[22]}
\begin{tabular}{|l|l|r|}
Model & Ref. & $B(\overline B\rightarrow X\ell\nu )$  [\%] \\
\tableline
ACCMM	& \cite{accmm}	& 10.56 $\pm$ 0.04 $\pm$ 0.22 \\
ISGW & \cite{isgw} & 10.26 $\pm$ 0.03 $\pm$ 0.22 \\
ISGW$^{**}$ & \cite{wang} & 10.96 $\pm$ 0.07 $\pm$ 0.22
\end{tabular}
\end{table}
\nopagebreak
\samepage

\begin{table}
\caption{The $b\rightarrow \bar cs$
$\rightarrow e^-$
background assumes $B(\overline B\rightarrow \overline DX) = 10\%$,
$B(\overline D^0 \rightarrow Xe^- \bar\nu ) =$ 6.64\%,
$\tau (D^+) / \tau (D^0 )$ = 2.55,
and
$B(D^{*-} \rightarrow\overline D^0 \pi^-)$ = 68.1\%.
The quantities $r$ and $w(s)$ are defined in the text.}
\begin{tabular}{|l|l|l|}
$r$ & $B(b\rightarrow \bar
cs\rightarrow e^- )|_{\overline B\rightarrow D\overline D \;
\overline K X} (\times 10^{-3})$
& $w(s = 0.2)$ \\
\tableline
\tableline
0 & 8.28 & 0.89 \\
\tableline
1/6 & 8.78 & 0.90 \\
\tableline
1/3 & 9.16 & 0.91  \\
\tableline
1 & 10.0 & 0.92

\end{tabular}
\end{table}

\begin{table} \caption{Classification of Semileptonic $B$ decays}
\begin{tabular}{|l|l|l|l|}
Class & Process & Branching Ratio (in \%) & Remark \\
\tableline
\tableline
$1^+$ & $\overline B^0 \rightarrow D^{*+} \ell^- \nu$ &
$(4.49 \pm 0.50)\;\left[\frac{4.01\%}{B(D^0 \rightarrow K^-
\pi^+)}\right]$ & Ref.~\cite{burchatr,barish} \\
\tableline
$1^0$ & $B^- \rightarrow D^{*o}\ell^- \nu$ & $(5.34 \pm 0.80)
\;\left[\frac{4.01\%}{B(D^0 \rightarrow K^- \pi^+)}\right]$ &
Ref.~\cite{burchatr} \\
\tableline
2 & $\overline B^0\rightarrow D^+ \ell\nu$ & $(1.69 \pm 0.36)
\;\left[\frac{4.01\%}{B(D^0 \rightarrow K^- \pi^+)}\right]$ &
missing mass technique of Ref.~\cite{kutschke} \\
\tableline
3 & $\overline B\rightarrow D^{**} \ell\nu$ & $\frac{\left(\rho^2 -
\frac{1}{4}\right)}{0.5} \;(0.8 \pm 0.4)$ & $D^{**}$ denotes the
narrow and
\\
& & & broad $p$ wave states, BR reflects \\
& & & Bjorken-Isgur-Wise sum rule \cite{sumrule} \\
\tableline
4 & $\overline B\rightarrow D^{r'} \ell\nu$ & 0.1 - 0.2 & $D^{r'}$
denotes
all
resonances beyond \\
& & & the $p$ wave states \cite{veselio,isgw2}. \\
\tableline
5 & $\overline B\rightarrow [X_c ]_{non-res} \ell\nu$ & 0.5 - 1.0 &
Nonresonant BR \cite{isgw2,veselio,yang,goityr,adgr} \\
\tableline
$1^++2-5$ & $\Sigma$ exclusives & 8.2 $\pm$ 0.9 & with $B(D^0
\rightarrow K^-
\pi^+) = (3.91 \pm 0.19) \%$ \\
\tableline
$1^0 +2-5$ & $\Sigma$ exclusives & 9.1 $\pm$ 1.1 & with $B(D^0
\rightarrow K^-
\pi^+) = (3.91 \pm 0.19) \%$ \\
\tableline
 $t$ & $\overline B\rightarrow X  \ell\nu$ & 10.49 $\pm$ 0.46
& Published inclusive semileptonic \\
& & & BR from ``model independent" \\
& & & dilepton analysis \cite{cleoslincl} \\

\end{tabular}
\end{table}


\clearpage


\begin{thebibliography}{999}


\bibitem{lepew96}
D. Abbaneo et al. (The LEP electroweak working group and the SLD
Heavy Flavor Group), A Combination of Preliminary LEP and SLD
Electroweak Measurements and Constraints on the Standard Model,
LEPEWWG/96-01, SLD Physics Note 47.

\bibitem{graziani}
E. Graziani (Delphi Collaboration), talk given at Fermilab, May 9,
1996.

\bibitem{charlton}
D. Charlton (OPAL collaboration), talk presented at the University of
Chicago, October 1995;
P. Antilogus et al. (LEP electroweak working group), CERN preprint,
CERN-PPE/95-172.

\bibitem{opalrc}
G. Alexander et al. (Opal Collaboration), CERN Report,
CERN-PPE/96-51, April 1996.

\bibitem{nchistory}
T.E. Browder, K. Honscheid, and S. Playfer, in {\it $B$ decays},
edited by S. Stone, 2nd edition (World Scientific, Singapore, 1994),
p. 158.

\bibitem{drellp}
J.P. Alexander et al. (CLEO collaboration),  CLEO conference report,
CLEO CONF 95-30.

\bibitem{cleod0}
D.S. Akerib et al. (CLEO collaboration), Phys.\ Rev.\ Lett.\ {\bf 71}, 3070 (1993) measures
$B(D^0 \rightarrow K^- \pi^+\;(\gamma) )= (3.95 \pm 0.19)\% \;$.
But because the detected $D^0 \rightarrow K^- \pi^+$ decays have an
effective cutoff on the radiative photons, CLEO uses for them
$(3.91 \pm 0.19)\% .$

\bibitem{pdg}
Particle Data Group, L. Montanet et al., Phys.\ Rev.\ {\bf D50}, 1173 (1994).

\bibitem{cleorp}
R. Balest et al. (CLEO Collaboration), Phys.\ Rev.\ Lett.\ {\bf 72}, 2328 (1994).

\bibitem{cleods}
M. Artuso et al. (CLEO Collaboration), Cornell report, CLNS-95-1387,
Jan. 1996.

\newpage

\bibitem{ncbrowder}
T.E. Browder, K. Honscheid and D. Pedrini [University of Hawaii
report,
UH-511-848-96, to appear in the 1996 Annual Review of Nuclear and
Particle Science] obtain from CLEO data the larger value
$$
n_c = 1.18 \pm 0.06 \;\;\;\;,
$$
for a number of reasons. First of all they have
combined the CLEO~\cite{cleod0}, ARGUS~\cite{argusd0}
and ALEPH~\cite{alephd0} measurements to
obtain
$$
B(D^0 \rightarrow K^- \pi^+ )= (3.76 \pm 0.15)\% \;.
$$
They then used the much larger measured inclusive $\Xi_c$ yield in $B$ decays
[where an error that propagates through the semileptonic $\Xi_c$ BR's
has not yet been corrected, see Ref.~\cite{jain}]. They also used
a factor of two  larger $Y_{\Lambda_c}$ than our central estimate
[because they used older CLEO 1.5 data and a smaller $B(\Lambda_c \to
p K \pi)$]. Finally, they have used a somewhat smaller measured
$B(\overline B\rightarrow (c\bar c) X)$.

\bibitem{argusd0}
H. Albrecht et al. (ARGUS collaboration), Phys.\ Lett.\ {\bf B340}, 125 (1994).

\bibitem{alephd0}
D. Decamp et al. (ALEPH collaboration), Phys.\ Lett.\ {\bf B266}, 218 (1991).

\bibitem{jain}
We thank Vivek Jain for  providing us with a corrected update of the
$\Xi_c$ production fractions in $B$ decays as measured by CLEO.

\bibitem{glasgowbaryon}
D. Cinabro et al. (CLEO Collaboration), Cornell report, CLEO CONF
94-8, 1994.

\newpage

\bibitem{shipsey}
T. Bergfeld et al. (CLEO Collaboration), Phys.\ Lett.\ {\bf B323}, 219 (1994).

\bibitem{moriond}
Y. Kwon (CLEO Collaboration), seminar presented at Moriond, March
1996.

\bibitem{bdy}
G. Buchalla, I. Dunietz, and H. Yamamoto, Phys.\ Lett.\ {\bf B364}, 188 (1995).

\bibitem{appel}
We thank Jeff Appel for pointing this fact out to us.

\bibitem{distw}
I. Dunietz, J. Incandela, R. Snider, K. Tesima, and I. Watanabe,
Fermilab report, FERMILAB-PUB-96/26-T, in progress.

\bibitem{burchatr}
J.D. Richman and P.R. Burchat, Rev. Mod. Phys. {\bf 67}, 893 (1995).

\bibitem{wang}
R. Wang (CLEO collaboration), Ph. D. Thesis, University of Minnesota
report, Dec. 1994.

\bibitem{cleoslincl}
B. Barish et al. (CLEO collaboration), Phys.\ Rev.\ Lett.\ {\bf 76}, 1570 (1996).

\bibitem{accmm}
G. Altarelli, N. Cabibbo, G. Corbo, L. Maiani, G. Martinelli, Nucl.\ Phys.\ {\bf B208}, 365 (1982).

\bibitem{isgw}  N. Isgur, D. Scora, B. Grinstein, and M.B. Wise,
Phys.\ Rev.\ {\bf D39}, 799 (1989).

\bibitem{isgw2}
D. Scora and N. Isgur, Phys.\ Rev.\ {\bf D52}, 2783 (1995).

\bibitem{argusdilepton}
H. Albrecht et al. (ARGUS Collaboration), Phys.\ Lett.\ {\bf B318}, 397 (1993).

\bibitem{alephslincl}
ALEPH Collaboration, Contribution to the IECHEP, EPS-HEP Brussels,
July-August 1995,
EPS95 Ref. eps0404.

\newpage

\bibitem{henderson}
S. Henderson (CLEO Collaboration), Phys.\ Rev.\ {\bf D45}, 2212
(1992).

\bibitem{browderh95}
T.E. Browder and K. Honscheid, Prog.\ Part.\ Nucl.\ Phys. {\bf 35}, 81 (1995).

\bibitem{neubert}
M. Neubert, CERN report, CERN-TH-96-055 [hep-ph/9604412], April 1996.

\bibitem{mixingwrong}
C. Albajar et al. (UA1 Collaboration), Phys.\ Lett.\ {\bf B262}, 171
(1991);
F. Abe et al. (CDF Collaboration), Phys.\ Rev.\ Lett.\ {\bf 67}, 3351
(1991);
K. Johns et al. (D0 Collaboration), Proceedings of the 5th
International Symposium on Heavy Flavor Physics, Montreal, Canada,
July 1993;
D. Decamp et al. (ALEPH Collaboration), Phys.\ Lett.\ {\bf B258}, 236
(1991);
B. Adeva et al. (L3 Collaboration), Phys.\ Lett.\ {\bf B288}, 395
(1992);
P. Abreu et al. (DELPHI Collaboration), Phys.\ Lett.\ {\bf B301}, 145
(1993);
P.D. Acton et al. (OPAL Collaboration), Phys.\ Lett.\ {\bf B276}, 379
(1992);
D. Buskulic et al. (ALEPH Collaboration), Z. Phys.\ {\bf C62}, 179
(1994);
CDF collaboration, CONTRIBUTED PAPER TO THE XVII INTL. SYMP. ON
LEPTON-PHOTON INTERACTIONS, BEIJING, CHINA, AUGUST 1995,
FERMILAB-CONF-95/230-E;
I. Dunietz, Fermilab report, FERMILAB-PUB-94-163-T [hep-ph/9409355],
Sep. 1994, unpublished.

\bibitem{sumrule}
J.D. Bjorken, invited talk given at Les Rencontre de la Valle
d'Aoste,
La Thuile, Italy, March 18-24, 1990;
N. Isgur and M.B. Wise, Phys.\ Rev.\ {\bf D43}, 819 (1991);
J.D. Bjorken, I. Dunietz, and J. Taron, Nucl.\ Phys.\ {\bf B371}, 111 (1992).

\bibitem{fwd}
A.F. Falk, M.B. Wise, and I. Dunietz, Phys.\ Rev.\ {\bf D51}, 1183 (1995).

\newpage

\bibitem{bsbsbar}
I. Dunietz, Phys.\ Rev.\ {\bf D52}, 3048 (1995).

\bibitem{bucbars}
The $b\rightarrow u\bar cs^\prime$ transitions were neglected at
present. They can be straightforwardly incorporated in future high precision
measurements.

\bibitem{bagan}
E. Bagan, P. Ball, V.M. Braun, P. Gosdzinsky, Phys.\ Lett.\ {\bf B342}, 362
(1995);
[Erratum to appear in Phys.\ Lett.\ {\bf B}];
E. Bagan, P. Ball, B. Fiol, P. Gosdzinsky, Phys.\ Lett.\ {\bf B351}, 546
(1995).

\bibitem{voloshin}
M.B. Voloshin, Phys.\ Rev.\ {\bf D51}, 3948 (1995).

\bibitem{baganrud}
E. Bagan, P. Ball, V.M. Braun, P. Gosdzinsky, Nucl.\ Phys.\ {\bf B432}, 3
(1994).

\bibitem{alephtau}
D. Buskulic et al. (ALEPH Collaboration), Phys.\ Lett.\ {\bf B343}, 444
(1995).

\bibitem{flnntau}
A.F. Falk, Z. Ligeti, M. Neubert, Y. Nir, Phys.\ Lett.\ {\bf B326}, 145
(1994).

\bibitem{voloshinqcd}
M. Beneke and G. Buchalla, private communication.  Arguments for large higher order
QCD corrections in $b\rightarrow c\bar cs^\prime$ transitions are given in M.B. Voloshin, TPI-MINN-96-1-T [hep-ph/9602256], Feb. 1996; and
M. Lu, M. Luke, M.J. Savage, B.H. Smith, University of Toronto report, UTPT-96-07 [hep-ph/9605406], May 1996.

\bibitem{neuberts}
M. Neubert and C.T. Sachrajda, CERN report, CERN-TH-96-19 [hep-ph/9603202], March
1996.

\bibitem{bbd}
M. Beneke, G. Buchalla, and I. Dunietz, SLAC report, SLAC-PUB-7165
[hep-ph/9605259], May 1996.

\bibitem{delphirc}
D. Bloch et al. (DELPHI Collaboration), DELPHI report, DELPHI 96-30 PHYS 604,
April 1996.

\bibitem{behnke}
T. Behnke (OPAL Collaboration), private communication.

\newpage

\bibitem{pvalue}
We choose here $p=0.10 \pm 0.05$ as motivated by the measured primary
production
fraction of $D_s$ mesons versus $D+D_s$ mesons. The larger $p$ used
in the Appendix is meant to demonstrate that even then
$Y_{\Xi_c}/Y_{\Lambda_c}$ is much smaller than currently believed.

\bibitem{nokstar}
All the arguments in this Section hold when the primary $\overline K$
is replaced by $\overline K^*$.  However, primary
$\overline K^*$ may not be copiously produced as briefly mentioned in
footnote 2 of Ref.~\cite{bdy}.

\bibitem{lipkinsanda}
H.J. Lipkin and A.I. Sanda, Phys.\ Lett.\ {\bf 201B}, 541 (1988).

\bibitem{june95}
I. Dunietz, Seminar given at CLEO, June 1995.

\bibitem{bjorkencolortransparency}
J.D. Bjorken, Nucl.\ Phys.\ {\bf B} (Proc.\ Suppl.) {\bf 11}
(1989) 325; {\bf SLAC-PUB-5389} (1990), published in Proc.\ of the
SLAC Summer Institute 1990, p.\ 167.

\newpage

\bibitem{bsw}
D. Fakirov and B. Stech, Nucl.\ Phys.\ {\bf B133}
(1978) 315;
L.L. Chau, Phys.\ Rep.\ {\bf B95} (1983) 1;
M. Bauer, B. Stech and M. Wirbel, Z. Phys.\ {\bf C34}
(1987) 103.

\bibitem{hqet}
For a review see, for instance, M. Neubert, Phys.\ Rep.\ {\bf 245}, 259 (1994).

\bibitem{bcincl}
To model the inclusive $\overline B\rightarrow X_c +\overline D^{(*)}
\overline K$ processes, we recommend to model the $\overline
B\rightarrow X_c$
transition via the underlying $V-A $ quark current $b\rightarrow c$,
as was done in a different context by W.F. Palmer and B. Stech,
Phys.\ Rev.\ {\bf D48}, 4174 (1993).

\bibitem{veselid}
S. Veseli and I. Dunietz, work in progress.

\bibitem{eichten}
E. Eichten, private communication.

\bibitem{distinguish}
I. Dunietz, Fermilab report, FERMILAB-PUB-94-163-T [hep-ph/9409355],
Sep. 1994, unpublished.

\bibitem{bigistone}
I. Bigi, B. Blok, M. Shifman, N. Uraltsev and A. Vainshtein, in
{\it $B$ Decays}, edited by S. Stone, 2nd edition (World Scientific,
Singapore,
1994), p.\ 132.

\newpage

\bibitem{schune}
M.-H. Schune, talk given at Moriond, March 1996.

\bibitem{wangthank}
We thank Roy Wang for many useful discussions.

\bibitem{neglectccdbkgd}
Because the Cabibbo suppressed transition $b\rightarrow c\bar cd$ is
below the presently available sensitivity, this subsection ignores it
for simplicity. It is possible to take it into account in future high
precision measurements.

\bibitem{chomenary}
X. Fu et al. (CLEO collaboration), CLEO report, CLEO CONF 95-11.

\bibitem{freyberger}
Y. Kubota et al. (CLEO collaboration), CLEO report, CLEO 95-18.

\bibitem{crawford}
G. Crawford et al. (CLEO collaboration), Phys.\ Rev.\ {\bf D45}, 752 (1992).

\bibitem{argusbaryon}
H. Albrecht et al., (ARGUS Collaboration) Z.\ Phys.\
{\bf C56}, 1 (1992).

\bibitem{dssubtract}
Where we understand that the small fraction of $e^+$ from right-charm
$\overline B\rightarrow (D^+_s, \; baryon_c) \rightarrow e^+$
has been subtracted from the raw data.

\newpage

\bibitem{cest} To estimate $c$, we note the following:
$$B(\overline B_d \rightarrow e^+ X)=B(\overline B_d \rightarrow DX)
B(D^0
\rightarrow e^+ X) \left[z_o +\left(1-z_o \right)\tau\right] \;,$$
$$B(B^- \rightarrow e^+ X)=B(B^- \rightarrow DX) B(D^0 \rightarrow
e^+ X)
\left[p_o +\left(1-p_o \right)\tau\right]\;,$$
where $D$ denotes the combined regular charm $D^0$ and $D^+$, and
$$z_o \equiv\frac{B(\overline B_d \rightarrow D^0 X)}{B(\overline B_d
\rightarrow
DX)} \;, \;\;p_o \equiv\frac{B(B^-\rightarrow D^0 X)}{B(B^-
\rightarrow
DX)} \;,$$
$$\tau\equiv\frac{B(D^+ \rightarrow e^+ X)}{B(D^0 \rightarrow e^+ X)}
\approx
\frac{\tau_{D^+}}{\tau_{D^0}} = 2.55 \;.$$
The parameter $c$ is thus given by
$$c=\frac{B(\overline B_d \rightarrow DX)}{B(B^- \rightarrow DX)}
\;\;\frac{\left[z_o +\left(1-z_o \right)\tau\right]}{\left[p_o
+\left(1-p_o
\right)\tau\right]}\;.$$
Our unsophisticated estimate chooses $B(\overline B_d \rightarrow
DX)/B(
B^- \rightarrow DX) \approx 1$, and can be refined by a careful
investigation of the
${\cal O}(1/m^3_b )$ effects to the $b\rightarrow c\bar ud$
transitions.
Information on the $z_o$ and $p_o$ parameters is contained in the
measurements
\cite{moriond} of
$$B(\overline B\rightarrow D^0 \ell^- X) \;\;{\rm and}\;\;
B(\overline
B\rightarrow D^+ \ell^- X)$$
and applying the factorization assumption. We are intrigued by what CLEO
obtains for $z_o$ and $p_o$ via this method.
We employed a cruder method, where we assumed factorization and that
the $b\rightarrow c$ transition predominantly gives rise to the lowest
lying $s$ wave states $D$ and $D^*$. [By observing $B(B^- \rightarrow D_1
(2420)^0 \pi^-
)\approx 0.16\;\%,$ CLEO demonstrated that our estimate has to be
refined. Our estimate has to be modified in more than one way.]
So our crude model predicts $p_o \approx 1$. As for $z_o$, we take
the ratio
$\Gamma (\overline B_d \rightarrow D^{*+}X) /\Gamma (\overline B_d
\rightarrow
D^+_{dir} X)$ to be equal to 3, motivated by
$\Gamma (\overline
B\rightarrow D^* \ell \bar\nu )/\Gamma (\overline B\rightarrow
D\ell\bar\nu )
\approx 3.$  We thus obtain $z_o = 0.51,$ which yields $c=1.76$.

\newpage

\bibitem{deltagamma}
J.S. Hagelin, Nucl.\ Phys.\ {\bf B193} (1981) 123;
E. Franco, M. Lusignoli and A. Pugliese, {\it ibid} {\bf B194} (1982)
403; L.L Chau, W.-Y. Keung and M. D. Tran, Phys.\ Rev.\ {\bf D27}
(1983) 2145; L.L Chau, Phys.\ Rep.\ {\bf 95} (1983) 1; A.J. Buras,
W. Slominski and H. Steger, Nucl.\ Phys.\ {\bf B245} (1984) 369;
M.B. Voloshin, N.G. Uraltsev, V.A. Khoze and M.A. Shifman, Yad.\
Fiz.\
{\bf 46} (1987) 181 [Sov.\ J. Nucl.\ Phys.\ {\bf 46} (1987) 112];
A. Datta, E.A. Paschos and U. T\"urke, Phys.\ Lett.\ {\bf B196}
(1987) 382;
M. Lusignoli, Z. Phys.\ {\bf C41} (1989) 645; R. Aleksan, A. Le
Yaouanc,
L. Oliver, O. P\`ene and Y.-C. Raynal, Phys.\ Lett.\ {\bf B316}
(1993) 567;
I. Bigi, B. Blok, M. Shifman, N. Uraltsev and A. Vainshtein, in
{\it B Decays}, edited by S. Stone, 2nd edition (World Scientific,
Singapore,
1994), p.\ 132 and references therein.

\bibitem{tenchini}
R. Tenchini (ALEPH Collaboration), private communication.

\bibitem{barish}
B. Barish et al. (CLEO Collaboration), Phys.\ Rev.\ {\bf D51}, 1014 (1995).

\bibitem{kutschke}
R. Kutschke (CLEO Collaboration), private communication.

\bibitem{arguspwave}
H. Albrecht (ARGUS Collaboration), Z.\ Phys.\ {\bf C57}, 533 (1993).

\bibitem{opalpwave}
R. Akers et al. (OPAL Collaboration), Z.\ Phys.\ {\bf C69}, 57 (1995).

\newpage

\bibitem{alephpwave}
D. Buskulic et al. (ALEPH Collaboration), Phys.\ Lett.\ {\bf B345}, 103 (1994)

\bibitem{bellerive}
A. Bellerive (CLEO collaboration), talk presented at the APS meeting, Indianapolis, Indiana, May 1996.

\bibitem{isgurprivate}
N. Isgur, private communication.

\bibitem{veselio}
S. Veseli and M.G. Olsson, MADPH-96-924 [hep-ph/9601307], Jan. 1996.

\bibitem{yang}
T.-M. Yan, H.-Y. Cheng, C.-Y. Cheung, G.-L. Lin, Y.C. Lin, and H.-L.
Yu, Phys.\ Rev.\ {\bf D46}, 1148 (1992).

\bibitem{goityr}
J.L. Goity and W. Roberts, Phys.\ Rev.\ {\bf D51}, 3459 (1995).

\bibitem{adgr}
D. Atwood, I. Dunietz, J.L. Goity and W. Roberts, in preparation.

\newpage

\bibitem{4spuzzle}
Evidently any additional experimental information regarding this
puzzle is desirable.
To that effect, consider that one $B$ is known to be charged. At an
$\Upsilon
(4S)$ experiment, the other $B$ is then also charged and becomes the
focus of our study. If this other $B$ is seen to decay semileptonically as
$$
B^- \rightarrow D^+ X^- \ell^- \bar\nu \;,
$$
then this proves either higher resonance [beyond $D^{(*)}$] or
nonresonant charm production in
semileptonic $B$ decay.  P. Drell informed us that B. Gittelman
independently suggested this idea and found that, at present, CLEO is
statistics limited and is unable to perform those measurements.

\bibitem{tnocharm}
The contribution due to the tiny $B(b \to u \ell \overline \nu)$ must be
subtracted from $BR(t)$.  Because it is significantly smaller than the error
on $BR(t)$ we neglect it in the following.

\bibitem{tipton}
P.L. Tipton (CLEO Collaboration), Ph. D. Thesis, University of
Rochester report, UR 984, 1987;
M.S. Alam (CLEO collaboration), Phys.\ Rev.\ Lett.\ {\bf 58}, 1814 (1987).

\bibitem{kwon}
This idea is being independently pursued by the CLEO Collaboration. Y. Kwon
and E. Thorndike, private communication.

\bibitem{menaryrp}
S. Menary, CBX95-88, September, 1995, unpublished.

\bibitem{argusdstar}
H. Albrecht et al. (ARGUS Collaboration), Phys.\ Lett.\ {\bf B324}, 249 (1994).

\bibitem{yamamotod0kpi}
H. Yamamoto, private communication for the CLEO collaboration.

\bibitem{yamamotobaryon}
We thank H. Yamamoto for a discussion on this point.

\bibitem{thorndike}
We thank E. Thorndike for a discussion on this point.

\bibitem{dcfw}
I. Dunietz, P.S. Cooper, A.F. Falk, and M.B. Wise, Phys.\ Rev.\ Lett.\ {\bf 73},
1075 (1994).

\newpage

\bibitem{dalitz}
Consider the $b\rightarrow cd\bar u$ transition
governed by the $V-A$ interaction.
The invariant mass of the $cd$ peaks at the highest possible values as
seen in a
Dalitz plot \cite{bdy}. Thus we expect the charmed
baryons initially produced
via $b\rightarrow c$ to be highly excited. In contrast, the $V-A$
nature of the
interaction favors smaller energies for the $\bar u$ antiquark in the
restframe of the decaying
$b$. Since the spectator antiquark $\bar q$ of the
$\overline B(\equiv
b\bar q )$ meson involves only a modest Fermi momentum, the invariant
mass of
the $\bar u\bar q$ system is also expected to be modest.

\bibitem{zoeller}
M.M. Zoeller (CLEO collaboration), Ph. D. Thesis, submitted to the
State University of New York, Albany, 1994.

\bibitem{tightlimit}The tight upper limit~\cite{crawford},
$$B(\overline B \to D^{*+} p \overline p X) < \; 0.4 \% \;(90\% \;
{\rm C.L.})\; ,$$
is not a problem for our scenario, because of flavor correlations.
The dominant charmed baryon producing process is governed by $b \to c
d \overline u$, and gives rise to highly excited baryons with flavor
$cdq$, where $q = u$ or $d$.  Strong decays of this highly excited
baryon can yield a $D^{(*)+}$, which is normally not correlated with
a proton, but rather
$$[cdq]^r \to D^{(*)+} \; [ddq] = D^{(*)+} (n, \Delta^{0,-}, ...)\; .$$
This naturally explains why no $D^{(*)+} p$ correlations from a
single $\overline B$ decay have been seen.  In contrast, we predict
larger $D^{(*)0} p$ correlations from single $\overline B$ decays,
via $$[cdq]^r \to D^{(*)0} \; p X\; .$$
Searches for these could be performed.

\bibitem{yamamoto}
H. Yamamoto (CLEO collaboration), private communication.

\bibitem{menary}
D. Gibaut et al. (CLEO Collaboration), Phys.\ Rev.\ {\bf D53}, 4734 (1996).

\newpage


\end{thebibliography}
\end{document}